\documentclass{article}

\usepackage{hyperref}
\usepackage[a4paper, top=25mm, left=10mm, right=10mm, bottom=30mm, headsep=10mm, footskip=12mm]{geometry}
\usepackage{authblk}

\usepackage[nointlimits]{amsmath}
\usepackage{siunitx}

\usepackage{caption}
\usepackage{graphicx}
\usepackage{subfigure}
\usepackage{xcolor}
\usepackage{lscape}

\usepackage[ruled,vlined]{algorithm2e}
\usepackage{algorithmic}

\usepackage{natbib}

\definecolor{mycolor_blue}{RGB}{66,124,161}
\definecolor{mycolor_grey}{RGB}{198,198,198}

\usepackage{tikz}
\usepackage{pgfplots}

\usetikzlibrary{patterns}

\usetikzlibrary{decorations.pathmorphing,arrows,intersections}

\pgfplotsset{
	kurze Legende/.style={
		legend image code/.code={
			\draw[##1,mark repeat=2,line width=0.6pt]
			plot coordinates {
				(0cm,0cm)
				(0.3cm,0cm)
			};
		}
	}
}

\pgfplotsset{
	compat = newest,
	scale only axis, 
	max space between ticks = 50pt,
	ticklabel style = {font=\footnotesize},
	legend style =  {font=\footnotesize},
	grid = major,
	grid style = {dotted},
	legend columns=1, 
	xtick pos=left,
	ytick pos=left
}

\newcommand{\longPicture}{
	\pgfplotsset{  
		width=0.8\textwidth,
		height=0.2\textwidth,
		ylabel style={text width=0.2\textwidth,align=center}
	}
}

\newcommand{\smallPicture}{
	\pgfplotsset{  
		width=0.35\textwidth,
		height=0.35\textwidth,
		ylabel style={text width=0.2\textwidth,align=center}
	}
}

\newcommand{\analytiSolutionPictures}{
	\pgfplotsset{  
		width=0.24\textwidth,
		height=0.25\textwidth,
		ylabel style={text width=0.2\textwidth,align=center}
	}
}

\pgfplotsset{select coords between index/.style 2 args={
		x filter/.code={
			\ifnum\coordindex<#1\fi
			\ifnum\coordindex>#2\fi
		}
}}

\definecolor{color1}{HTML}{0060AD} 
\definecolor{color2}{HTML}{FF4500} 
\definecolor{color3}{HTML}{FFA500} 
\definecolor{color4}{HTML}{006400} 
\definecolor{color5}{HTML}{9400D3} 
\definecolor{color6}{HTML}{800000} 
\definecolor{color7}{HTML}{000000} 
\definecolor{color8}{HTML}{0000FF} 
\definecolor{color9}{HTML}{FF0000} 
\definecolor{mycolor_blue}{RGB}{66,124,161}
\definecolor{mycolor_grey}{RGB}{198,198,198} 

\tikzstyle{line1} = [color=color7,semithick] 
\tikzstyle{line2} = [color=color2,densely dotted,semithick]
\tikzstyle{line3} = [color=color1,densely dashed,semithick]
\tikzstyle{line4} = [color=color5,dash dot,semithick]
\tikzstyle{line5} = [color=color4,dash dot dot,semithick]
\tikzstyle{line6} = [color=color6,semithick]

\tikzstyle{mark1} = [color=color7,mark=x,mark size=2pt,mark options=solid,semithick] 
\tikzstyle{mark2} = [color=color2,mark=o,mark size=2pt,mark options=solid,semithick]
\tikzstyle{mark3} = [color=color1,mark=*,mark size=2pt,mark options=solid,semithick]
\tikzstyle{mark4} = [color=color5,mark=triangle,mark size=2pt,mark options=solid,semithick]
\tikzstyle{mark5} = [color=color4,mark=square,mark size=2pt,mark options=solid,semithick]

\begin{document}

\title{Continuous Adjoint Complement to the Blasius Equation}

\author{Niklas K\"uhl\thanks{niklas.kuehl@tuhh.de}, Peter M. M\"uller and Thomas Rung}

\affil{Hamburg University of Technology, Institute for Fluid Dynamics and Ship Theory, Am Schwarzenberg-Campus 4, D-21075 Hamburg, Germany}

\maketitle

\begin{abstract}
The manuscript is concerned with a continuous adjoint complement to two-dimensional, incompressible, first-order boundary-layer equations for a  flat plate boundary-layer. The text is structured into three parts.
The first part demonstrates, that the  adjoint complement can be derived in two ways, either following a \textit{first simplify then derive} or a \textit{first derive and then simplify} strategy. The simplification step comprises the classical boundary-layer (b.-l.) approximation and the derivation step transfers the primal flow equation into a companion adjoint equation. 

The second part of the paper comprises the analyses of the coupled primal/adjoint b.-l. framework. This leads to similarity parameters, which turn the  Partial-Differential-Equation (PDE) problem into a boundary value problem described by a set of Ordinary-Differential-Equations (ODE) and support the formulation of an adjoint complement to the classical Blasius equation.
Opposite to the primal Blasius equation, its adjoint complement consists of two ODEs which can be simplified depending on the treatment of advection. It is shown, that the advective fluxes, which are frequently debated in the literature, vanish for the investigated self-similar b.l. flows. Differences between the primal and the adjoint Blasius framework are discussed against numerical solutions, and analytical expressions are derived for the adjoint b.-l. thickness, wall shear stress and subordinated skin friction and drag coefficients. The analysis also provides an analytical expression for the shape sensitivity to shear driven drag objectives.

The third part assesses the predictive agreement between the different Blasius solutions and numerical results for Navier-Stokes simulations of a flat plate b.-l. at Reynolds numbers between $10^3 \leq \mathrm{Re}_\mathrm{L} \leq 10^5$. It is seen, that the reversal of the inlet and outlet location and the direction of the flow, inherent to the adjoint formulation of convective kinematics, poses a challenge when investigating real finite length (finite Re-number) flat plate boundary layer problems. Efforts to bypass related issues are discussed.
\end{abstract}


\section{Introduction}

The goal of an adjoint analysis is commonly the efficient computation of derivative information of an integral objective functional with respect to a general control function. The approach has recently gained increasing popularity in local optimization strategies using computational fluid dynamics (CFD) 
\cite{springer2015adjoint, papoutsis2016continuous, thomas2016adjoint, kroger2018adjoint, kapellos2019unsteady}.
In continuous space, the dual or adjoint flow state can be interpreted as a co-state and always follows from the underlying primal PDE governed model, which describes the flow physics. The cost function definition typically provides necessary boundary conditions and/or source terms to close the formulation as well as a sound relation between the objective functional and the control.

The formulation of boundary conditions, an appropriate treatment of advective terms and the adequate discretization  are often not intuitively clear in a PDE-based, continuous adjoint (CA) framework \cite{soto2004computation, othmer2008continuous, othmer2014adjoint, springer2015adjoint, kroger2018adjoint}.
Therefore the understanding of adjoint flows as well as the development of numerical strategies clearly lags behind the primal progress, and has motivated the development of discrete adjoint (DA) approaches using automatic differentiation to synchronize the primal and dual development states, e.g. \cite{griewank1989automatic, griewank2000algorithm, nadarajah2000comparison, nadarajah2003discrete}. 
The DA approach passes over the adjoint PDE and directly bridges the discrete linearized primal flow into a consistent discrete dual approach, cf. a comprehensive discussion in Giles and Pierces \cite{giles1997adjoint,giles2000introduction} or the lecture series by Vassberg and Jameson \cite{vassberg2006aerodynamic, vassberg2006aerodynamic_II}. 
Despite the various merits and drawbacks of the DA vs. the CA method, the CA approach is  unique for its invaluable contribution to a physical understanding and will therefore be the method of choice in the present paper. 

We intent to contribute to understanding adjoint wall-bounded shear flows by deriving the adjoint complement to a first-order boundary-layer (b.-l.) framework. 
 The paper tries to answer the fundamental question, if there is an adjoint complement to the similarity transformation inspired by the Blasius solution for a laminar flat plate boundary-layer, 
 and examines different paths towards a generalized primal/adjoint similarity transformation.  
  With a view to practical implications, attention is given to the influence of the adjoint advection, frequently labeled  adjoint transpose convection (ATC). The ATC term is often debated for the invoked robustness and discretization issues \cite{elliott1997practical, anderson1999airfoil, soto2004computation, othmer2008continuous, peter2010numerical, kroger2018adjoint}, and its 
  appearance in analytical b.-l. formulations is of practical interest.

Starting from the non-linear Navier-Stokes (NS) equations, different routes to derive an adjoint b.-l. framework are conceivable. One could first simplify the primal NS equations using the classical b.-l. approximation for $Re\to \infty$ to formulate the b.-l. equations, subsequently linearize the b.-l. equations and finally derive their adjoint companion.  On the contrary, the NS equations could be linearized in an initial step and the adjoint NS companion can subsequently be derived, which is finally reduced to an adjoint b.-l. approximation for $Re \to \infty$.
In a sense, the two options remind of the above discussed options for obtaining the discrete adjoint from a derive-and-discretize (CA) or discretize-and-derive (DA) strategy. Both approaches, i.e. {\it simplify-and-derive} as well as {\it derive-and-simplify} will be performed for the first time in this paper, to verify their agreement and thus correctness. 
 A more desirable, compact form of the primal b.-l. equations employs a similarity transformation and reduces the b.-l. PDE system into a single ODE in line with the Blasius solution.
Although the similarity transformation 
is expected to be generalizable, an adjoint similarity transformation was -- to the best of our knowledge -- never examined directly.
The paper outlines a strategy to derive an adjoint complement to the Blasius equation from a general similarity analysis. Interestingly, the adjoint Blasius equations furnish evidence that the ATC term -- despite its formal occurrence -- remains identical to zero, a fact that is not obvious from the b.-l. PDEs. Supplementary we demonstrate, that the application of the adjoint Blasius equation to a drag objective functional displays strong similarities to a thermal b.-l. solution at negative unit Prandtl number.

Previous adjoint-based investigations into b.-l. type flows can be grouped into investigation with respect to (w.r.t.) either b.-l. receptivity \cite{airiau2002boundary, giannetti2006leading}, modal sensitivity \cite{giannetti2007structural, giannetti2010structural} or b.-l. stability \cite{herbert1997parabolized, hwang2006control} considering the perturbation-based excitation of Tollmien-Schlichting waves \cite{airiau2000non, airiau2003methodology} or the linearized Orr-Sommerfeld equation \cite{giannetti2006leading, bottaro2003effect}.
Hill \cite{hill1992theoretical,  hill1995adjoint} was the first who investigated the adjoint Orr-Sommerfeld operator and subsequently extended his study towards the adjoint to a Parabolic Stability Equation. The latter was enhanced by Pralits et al. \cite{pralits2002adjoint} to derive an adjoint b.-l. framework for the optimization of disturbance control -- that somehow differs to our access into the topic of adjoint wall-bounded flow.
A very good overview of adjoint applications to fluid-dynamic stability analysis is given in Luchini and Bottaro \cite{luchini2014adjoint}.
Unlike the previous research, the present paper derives the complete adjoint set of b.-l. equations and compares them with their primal companion. Hence, instead of transposing a few -- although particularly important -- primal operators, we offer an extensive adjoint-based insight and thus should sharpen the adjoint b.-l. understanding.
The  investigated b.-l. flow is admittedly simple. However, we still consider it of fundamental interest for engineering applications. Whilst findings 
are not provably valid for more complex flows, e.g. turbulent b.-l. or separated flows, they might still by indicative for attached boundary-layers and other virtually unidirectional shear flows \cite{schrader2011receptivity, tempelmann2012swept}. Of particular interest is the disappearance of the ATC term in the Blasius framework, which corresponds to an often employed heuristic simplification. The issue will also be discussed in the application section which compares different primal/adjoint NS formulations with results of a primal/adjoint Blasius framework.

The remainder of the paper is organised as follows: Section \ref{sec:primal_field_equations} is concerned with the derivation of the adjoint b.-l. equations. The \ref{sec:adjoint_blasius_equation}rd section presents a generalized similarity transformation to condense the primal/dual PDE systems into an ODE framework. Subsequently, we investigate the adjoint Blasius equation numerically in Sec. \ref{sec:numerical_blasius_approximation} as well as analytically in Sec. \ref{sec:continuous_blasius_investigation}. Section \ref{sec:application} compares b.-l. results with NS solutions for a zero pressure gradient (ZPG) flat plate flow, which are obtained from numerical simulations at Reynolds numbers between $10^3 \leq \mathrm{Re}_\mathrm{L} \leq 10^5$. The final section \ref{sec:conclusion} provides conclusions and outlines future research. Within the publication, Einstein’s summation convention is used for lower-case Latin subscripts. Vectors and tensors are defined with reference to Cartesian coordinates and dimensionless field quantities are consistently marked with an Asterisk.
\section{Primal and Adjoint Boundary-Layer Equations}
\label{sec:primal_field_equations}
The paper deals with incompressible fluids in steady-state that form a flat-plate momentum boundary layers exposed to zero pressure gradients. More generally, the velocity $v_\mathrm{i}$ and pressure $p$ follow from the steady incompressible NS equations 
\begin{alignat}{2}
R_\mathrm{i}:& v_\mathrm{k} \frac{\partial v_\mathrm{i}}{\partial x_\mathrm{k}} + \frac{\partial}{\partial x_\mathrm{k}} \left[ \frac{p}{\rho} \delta_\mathrm{ik} - 2 \nu S_\mathrm{ik} \right] &&= 0 \label{equ:primal_momentum} \\
Q:& -\frac{\partial v_\mathrm{k}}{\partial x_\mathrm{k}} &&= 0 \label{equ:primal_mass}.
\end{alignat}
where $\rho$, $\nu$, $S_\mathrm{ik} = 1/2 ( \partial v_\mathrm{i} / \partial x_\mathrm{k} +  \partial v_\mathrm{k} / \partial x_\mathrm{i} )$ and $\delta_\mathrm{ik}$ represent density, kinematic viscosity, the symmetric strain rate tensor as well as the Kronecker delta respectively.
Boundary conditions are given in Tab. (\ref{tab:bound_condi}). 

Two-dimensional equations for the momentum boundary layer in the $x_1$-$x_2$-plane can be derived in two consecutive steps. First, the governing Eqn. (\ref{equ:primal_momentum})-(\ref{equ:primal_mass}) are non-dimensionalized with  the reference quantities given in Tab. \ref{tab:reference_values}. An exemplary relationship between a dimensional term, his reference values and non-dimensional quantities marked with an asterisk reads $v_\mathrm{2} \partial v_\mathrm{1} / \partial x_\mathrm{2}  = ( V_\mathrm{2} V_\mathrm{1} / \delta ) \,( v_\mathrm{2}^* \partial v_\mathrm{1}^* / \partial x_\mathrm{2}^*)$.
Subsequently, a scaling analysis is performed, where it is assumed that the spatial extent in streamwise $x_\mathrm{1}$-direction is significantly larger than the extent in the direction of the wall normal  $x_\mathrm{2}$, viz. $L \gg \delta$.
The assumption is in line with a flat plate of semi-infinite length $L$ ($\to \infty$) in the $x_\mathrm{1}$ direction (cf. Fig, \ref{fig:plate_flow}).
Thus, the continuity Eqn. (\ref{equ:primal_mass}) reveals $V_\mathrm{2} \propto V_\mathrm{1} \delta / L$.
The scaling analysis is performed in App. \ref{app:primal_simplification} using the reference data from Tab. (\ref{tab:reference_values}). Here, only the resulting b.-l. equations are given in residual notation, i.e. 
\begin{alignat}{3}
R_\mathrm{1}^\mathrm{BL}:& v_\mathrm{1} \frac{\partial v_\mathrm{1}}{\partial x_\mathrm{1}} + v_\mathrm{2} \frac{\partial v_\mathrm{1}}{\partial x_\mathrm{2}} + \frac{1}{\rho} \frac{\partial p}{\partial x_\mathrm{1}}  - \nu \frac{\partial^2 v_\mathrm{1}}{\partial {x_\mathrm{2}}^2}  &= 0 \label{equ:primal_bounday_layer_mome1_first} \; , \\
R_\mathrm{2}^\mathrm{BL}:& \frac{1}{\rho} \frac{\partial p}{\partial x_\mathrm{2}}  &= 0  \label{equ:primal_bounday_layer_mome2_first} \; , \\
Q^\mathrm{BL}:&  - \left( \frac{\partial v_\mathrm{1}}{\partial x_\mathrm{1}} + \frac{\partial v_\mathrm{2}}{\partial x_\mathrm{2}} \right) &= 0  \label{equ:primal_bounday_layer_mass1_first}.
\end{alignat}

\begin{table}
\centering
\begin{tabular}{|c||c|c||c|c|}
\hline
boundary type & $v_\mathrm{i}$  & $p$ & $\hat{v}_\mathrm{i}$  & $\hat{p}$\\
\hline
\hline
inlet & $v_\mathrm{i} = v_\mathrm{i}^\mathrm{in}$ & $\frac{\partial p}{\partial n} = 0$ & $\hat{v}_\mathrm{i} n_\mathrm{i} = - \rho \frac{\partial j_\mathrm{\Gamma}}{\partial p} $ & $\frac{\partial \hat{p}}{\partial n} = 0$  \\
\hline
outlet & $\frac{\partial v_\mathrm{i}}{\partial n} = 0$ &  $\frac{\partial p}{\partial n} = 0$  & $\hat{v}_\mathrm{i} n_\mathrm{i} = - \rho \frac{\partial j_\mathrm{\Gamma}}{\partial p}$ & $\hat{p} n_\mathrm{i} = v_\mathrm{k} \hat{v}_\mathrm{i} n_\mathrm{k} + 2 \mu S_\mathrm{ik} + \frac{\partial j_\mathrm{\Gamma}}{\partial v_\mathrm{i}}$    \\
\hline
wall & $v_\mathrm{i} = 0$ & $\frac{\partial p}{\partial n} = 0$ & $\hat{v}_\mathrm{i} n_\mathrm{i} = - \rho \frac{\partial j_\mathrm{\Gamma}}{\partial p}$ &  $\frac{\partial \hat{p}}{\partial n} = 0$\\
\hline
\end{tabular}
\caption{Boundary conditions for the primal and adjoint equations.}
\label{tab:bound_condi}
\end{table}

\begin{table}
\centering
\begin{tabular}{|c||c|c|c|c|c|}
\hline
primal quantity/operator & 
$v_\mathrm{1}$  & $v_\mathrm{2}$  & $p$ & $ \partial \phi / \partial x_\mathrm{1}$ & $ \partial \phi / \partial x_\mathrm{2}$ \\
\hline
reference value & $V_\mathrm{1}$ & $V_\mathrm{2}$ & $P$ & $L$  & $\delta$\\
\hline
\hline
adjoint quantity/operator & 
$ \hat{v}_\mathrm{1}$  & $\hat{v}_\mathrm{2}$  & $\hat{p}$ & $ \partial \hat{\phi} / \partial x_\mathrm{1}$ & $ \partial \hat{\phi} / \partial x_\mathrm{2}$ \\
\hline
reference value & $\hat{V}_\mathrm{1}$ & $\hat{V}_\mathrm{2}$ & $\hat{P}$ & $L$  & $\hat{\delta}$ \\
\hline
\end{tabular}
\caption{Reference quantities of the 2D governing equations.}
\label{tab:reference_values}
\end{table}

As noted above, the corresponding adjoint b.-l. equations can be derived in two ways, following either a \emph{derive-and-simplify} or a \emph{simplify-and-derive} strategy. In both cases the derivation step starts with the definition of a surface [volume] based objective functional $j_\mathrm{\Gamma}$ [$j_\mathrm{\Omega}$] that allows for the construction of an augmented objective, frequently labeled as Lagrangian $L$. 
 Following the \emph{derive-and-simplify} route, one obtains 
\begin{align}
L &= \int_\mathrm{\Gamma} j_\mathrm{\Gamma} \mathrm{d} \Gamma + \int_{\mathrm{\Gamma}_\mathrm{in}} \hat{v_\mathrm{i}} \left[ v_\mathrm{i} - n_\mathrm{i} v_\mathrm{i}^\mathrm{in} \right] + \hat{p} \frac{\partial p}{\partial n} \mathrm{d} \Gamma \nonumber \\
& + \int_{\mathrm{\Gamma}_\mathrm{out}} \hat{v_\mathrm{i}} \left[ \frac{\partial v_\mathrm{i}}{\partial n} \right] + \hat{p} \frac{\partial p}{\partial n} \mathrm{d} \Gamma + \int_{\mathrm{\Gamma}_\mathrm{wall}} \hat{v_\mathrm{i}} v_\mathrm{i} + \hat{p} \frac{\partial p}{\partial n} \mathrm{d} \Gamma \nonumber \\
& + \int_{\Omega_\mathrm{O}} j_\mathrm{\Omega} \mathrm{d} \Omega + \int_{\Omega}  \hat{v}_\mathrm{i} R_\mathrm{i} + \hat{p} \, Q \mathrm{d} \Omega,
\end{align}
where the second volume integral accounts for the field Eqns. (\ref{equ:primal_momentum})-(\ref{equ:primal_mass}) and the boundary integrals consider the corresponding boundary conditions from Tab. (\ref{tab:bound_condi}), $\Gamma = \Gamma_\mathrm{in} \cap \Gamma_\mathrm{out} \cap \Gamma_\mathrm{wall}$.
Mind that the habitat of an objective does not necessarily coincide with the complete surface or volume, which is why we introduce an additional subscript, viz. $\Gamma_\mathrm{O} \cap \Gamma$ and $\Omega_\mathrm{O} \cap \Omega$ . The local objectives $j_\mathrm{\Gamma}$ [$j_\mathrm{\Omega}$] disappear on $\Gamma \backslash \Gamma_\mathrm{O}$ [in $\Omega \backslash \Omega_\mathrm{O}$].
The Lagrangian multipliers $\hat{v}_\mathrm{i}$ and $\hat{p}$ are frequently labeled as dual or adjoint velocity and pressure. The dimensions of $\hat{v}_\mathrm{i}$ and $\hat{p}$ depend on the underlying objective, e.g. $[\hat{v}_\mathrm{i}] = [J]/( [R_\mathrm{i}] \, m^3)$ and $[\hat{p}] = [J] / ([Q] \, m^3)$ where $[J] = [j_\mathrm{\Omega}] \, m^3$ ($[J] = [j_\mathrm{\Gamma}] \, m^2$) represents the units of the integral volume-based (surface-based) objective.
First order optimality conditions force a vanishing derivative of the Lagrangian in all dependent directions
($\delta_{\mathrm{v}_\mathrm{i}} L \cdot \delta v_\mathrm{i} = 0 \ \forall  \delta v_\mathrm{i}$, $\delta_\mathrm{p} L \cdot \delta p = 0 \ \forall \delta p$). The related derivatives read:
\begin{align}
\delta_\mathrm{v_\mathrm{i}} L \cdot \delta v_\mathrm{i} &= 
\int_{\mathrm{\Gamma}_\mathrm{in}} \hat{v_\mathrm{i}} (\delta v_\mathrm{i}) \mathrm{d} \Gamma  
+ \int_{\mathrm{\Gamma}_\mathrm{out}} \hat{v_\mathrm{i}} \frac{\partial (\delta v_\mathrm{i})}{\partial n} \mathrm{d} \Gamma 
+ \int_{\mathrm{\Gamma}_\mathrm{wall}} \hat{v_\mathrm{i}} (\delta v_\mathrm{i}) \mathrm{d} \Gamma  \nonumber \\
&\hspace{5mm} + \int_{\mathrm{\Gamma}}\delta v_\mathrm{i} \left[ \frac{\partial j_\mathrm{\Gamma}}{\partial v_\mathrm{i}} + v_\mathrm{k} \hat{v}_\mathrm{i} n_\mathrm{k} + 2 \nu \hat{S}_\mathrm{ik} n_\mathrm{k} - \hat{p} n_\mathrm{i} \right] - (\delta S_\mathrm{ik}) \hat{v}_\mathrm{i} \nu \mathrm{d} \Gamma \nonumber \\
&\hspace{5mm} + \int_{\Omega} \delta v_\mathrm{i} \left[ \hat{v}_\mathrm{k} \frac{\partial v_\mathrm{k}}{\partial x_\mathrm{i}} - v_\mathrm{k} \frac{\partial \hat{v}_\mathrm{i}}{\partial x_\mathrm{k}} - 2 \nu \frac{\partial \hat{S}_\mathrm{ik}}{\partial x_\mathrm{k}} + \frac{\partial \hat{p} }{\partial x_\mathrm{i}} + \frac{\partial j_\mathrm{\Omega}}{\partial v_\mathrm{i}} \right] \mathrm{d} \Omega \nonumber \\
& \overset{!}{=} 0 \ \forall  \delta v_\mathrm{i} \label{equ:vel_der_vol}  \\
\delta_\mathrm{p} L \cdot \delta p &= 
\int_{\mathrm{\Gamma}_\mathrm{in}} \hat{p} \frac{\partial \delta p}{\partial n} \mathrm{d} \Gamma 
 + \int_{\mathrm{\Gamma}_\mathrm{out}} \hat{p} \frac{\partial \delta p}{\partial n} \mathrm{d} \Gamma 
 + \int_{\mathrm{\Gamma}_\mathrm{wall}} \hat{p} \frac{\partial \delta p}{\partial n} \mathrm{d} \Gamma \nonumber \\
&\hspace{5mm} + \int_\mathrm{\Gamma} \delta p \left[\frac{\partial j_\mathrm{\Gamma}}{\partial p} + \hat{v}_\mathrm{i} \frac{1}{\rho} n_\mathrm{i} \right] \mathrm{d} \Gamma \nonumber \\
&\hspace{5mm} + \int_{\Omega} \delta p \left[ -  \frac{\partial \hat{v}_\mathrm{i}}{\partial x_\mathrm{i}} \frac{1}{\rho} + \frac{\partial j_\mathrm{\Omega}}{\partial p} \right] \mathrm{d} \Omega \nonumber  \\ 
& \overset{!}{=} 0 \ \forall \delta p.  \label{equ:pres_der_vol}
\end{align}
Here, $\hat S_\mathrm{ik} = 1/2 ( \partial \hat{v}_\mathrm{i} / \partial x_\mathrm{k} +  \partial \hat{v}_\mathrm{k} / \partial x_\mathrm{i} )$ represents the adjoint strain rate tensor.
The adjoint field equations result from the field integrals in Eqns. (\ref{equ:vel_der_vol})-(\ref{equ:pres_der_vol}), hence  
\begin{alignat}{2}
\hat{R}_\mathrm{i}:& \hspace{1cm}  - v_\mathrm{k} \frac{\partial \hat{v}_\mathrm{i}}{\partial x_\mathrm{k}} + \hat{v}_\mathrm{k} \frac{\partial v_\mathrm{k}}{\partial x_\mathrm{i}} + \frac{\partial}{\partial x_\mathrm{k}} \left[ \hat{p} \delta_\mathrm{ik} - 2 \nu \hat{S}_\mathrm{ik} \right] &&= -\frac{\partial j_\mathrm{\Omega}}{\partial v_\mathrm{i}} \label{equ:adjoint_momentum} \; \\
\hat{Q}:& \hspace{5.7cm} -\frac{1}{\rho}\frac{\partial \hat{v}_\mathrm{k}}{\partial x_\mathrm{k}} &&= -\frac{\partial j_\mathrm{\Omega}}{\partial p} \label{equ:adjoint_mass} \; . 
\end{alignat}
The corresponding boundary conditions follow from the surface integrals in Eqn. (\ref{equ:vel_der_vol})-(\ref{equ:pres_der_vol}) and are summarized in Tab. (\ref{tab:bound_condi}). They depend on the underlying surface based objective functional $j_\mathrm{\Gamma}$ and we refer to \cite{kuhl2019decoupling,kuhl2020adjoint} for a more detailed discussion.

Mind that an additional cross coupling term, frequently labeled as ATC, arises in the adjoint system due to the non-linearity of the primal convective momentum transport -- e.g. the second term on the left hand side in (\ref{equ:adjoint_momentum}). The term might disappear 
for compressible flows, cf. Soto and L{\"o}hner \cite{soto2004computation}. Nevertheless, the ATC terms are  frequently also neglected in incompressible formulations due to 
the related impairment of the numerical robustness \cite{springer2015adjoint, othmer2008continuous}.  Some authors raise  a \emph{mathematical} argument based on the approximation order, viz. 
$\delta [\underline{v} \cdot (\underline{\nabla} \, \underline{v})] = \delta \underline{v} \cdot (\underline{\nabla} \, \underline{v}) + \underline{v} \cdot (\underline{\nabla} \, \delta \underline{v}) \approx \mathcal{O} (\delta \underline{v}) + \mathcal{O} (\underline{\nabla} \, \delta \underline{v})$ to justify the neglect of the ATC \cite{anderson1999airfoil, elliott1997practical, othmer2008continuous}. 
The present manuscript provides another 
indicator for the \emph{physically} justifiable neglect of this term.

Similar to the primal problem, the adjoint system (\ref{equ:adjoint_momentum})-(\ref{equ:adjoint_mass}) is analyzed w.r.t. the spatial scale/order of magnitude in a non-dimensional setting. The corresponding reference quantities follow from Tab. (\ref{tab:reference_values}).
We define a scaling of the $x_\mathrm{2}$-derivative of an adjoint quantity with $\hat{\delta}$, viz. $v_\mathrm{2} \partial \hat{v}_\mathrm{1} / \partial x_\mathrm{2}  = ( V_\mathrm{2} \hat{V}_\mathrm{1} / \hat{\delta} ) \,( v_\mathrm{2}^* \partial \hat{v}_\mathrm{1}^* / \partial x_\mathrm{2}^*)$.
The scaling analysis of Eqn. (\ref{equ:adjoint_momentum})-(\ref{equ:adjoint_mass}) is performed in App. \ref{app:adjoint_simplification} for an objective that does not depend on the primal pressure, e.g. $\partial j_\mathrm{\Omega} / \partial p = 0$ and yields:
\begin{alignat}{3}
\hat{R}_\mathrm{1}^\mathrm{BL}:& -v_\mathrm{1} \frac{\partial \hat{v}_\mathrm{1}}{\partial x_\mathrm{1}} - v_\mathrm{2} \frac{\partial \hat{v}_\mathrm{1}}{\partial x_\mathrm{2}} + \hat{v}_\mathrm{1} \frac{\partial v_\mathrm{1}}{\partial x_\mathrm{1}} + \frac{\partial \hat{p}}{\partial x_\mathrm{1}} - \nu \frac{\partial^2 \hat{v}_\mathrm{1}}{\partial {x_\mathrm{2}}^2} &&= -\frac{\partial j_\mathrm{\Omega}}{\partial v_\mathrm{1}} \label{equ:adjoint_bounday_layer_mome1_first}\\
 \hat{R}_\mathrm{2}^\mathrm{BL}:  & \; \hat{v}_\mathrm{1} \frac{\partial v_\mathrm{1}}{\partial x_\mathrm{2}} + \frac{\partial \hat{p}}{\partial x_\mathrm{2}} &&= -\frac{\partial j_\mathrm{\Omega}}{\partial v_\mathrm{2}} \label{equ:adjoint_bounday_layer_mome2_first}\\
 \hat{Q}^\mathrm{BL}:& - \left( \frac{1}{\rho} \frac{\partial \hat{v}_\mathrm{1}}{\partial x_\mathrm{1}} + \frac{1}{\rho}  \frac{\partial \hat{v}_\mathrm{2}}{\partial x_\mathrm{2}} \right) &&=0 \label{equ:adjoint_bounday_layer_mass_first} 
\end{alignat}

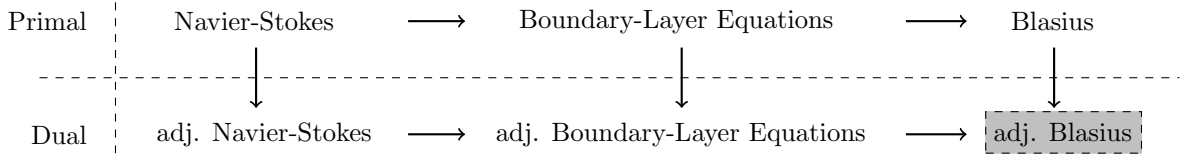
\begin{figure}
\centering
\begin{tikzpicture}

\draw[dashed] (-5,0) -- (10,0);
\draw[dashed] (-4,1) -- (-4,-1);

\draw (-4.25,+0.75) node [anchor=east] {Primal};
\draw (-4.25,-0.75) node [anchor=east] {Dual};

\draw (-1.00,+0.75) node [anchor=east] {Navier-Stokes};
\draw (-0.50,-0.75) node [anchor=east] {adj. Navier-Stokes};
\draw [->,black,thick] (-2.15,0.4) -- (-2.15,-0.4);

\draw [->,black,thick] (-0.15,0.75) -- (0.6,0.75);
\draw [->,black,thick] (-0.15,-0.75) -- (0.6,-0.75);
\draw (5.575,0.75) node [anchor=east]  {Boundary-Layer Equations};
\draw (6.00,-0.75) node [anchor=east] {adj. Boundary-Layer Equations};
\draw [->,black,thick] (3.45,0.4) -- (3.45,-0.4);

\draw [->,black,thick] (6.4,0.75) -- (7.2,0.75);
\draw [->,black,thick] (6.4,-0.75) -- (7.2,-0.75);
\draw (9.00,0.75) node [anchor=east]  {Blasius};
\draw (9.50,-0.75) node [anchor=east] {adj. Blasius};
\draw [->,black,thick] (8.35,0.4) -- (8.35,-0.4);

\filldraw[fill=black,opacity=0.25] (7.45,-1.0) -- (9.5,-1.0) -- (9.5,-0.45) -- (7.45,-0.45) -- (7.45,-1.0);
\draw[dashed] (7.45,-1.0) -- (9.5,-1.0) -- (9.5,-0.45) -- (7.45,-0.45) -- (7.45,-1.0);

\end{tikzpicture}
\caption{Schematic derivation-flow of adjoint counterparts to known primal simplifications for near-wall flow physics towards the desired adjoint Blasius equation (dark grey).}
\label{fig:derivation_overview}
\end{figure}

Alternatively 
Eqns. (\ref{equ:adjoint_bounday_layer_mome1_first})-(\ref{equ:adjoint_bounday_layer_mass_first}) can also be derived in a \textit{simplify-and-derive} approach (cf. Fig. \ref{fig:derivation_overview}). The latter starts with a Lagrangian based on the primal b.-l. equations  (\ref{equ:primal_bounday_layer_mome1_first})-(\ref{equ:primal_bounday_layer_mass1_first}) , viz.
\begin{align}
L &= \int_\mathrm{\Gamma} j_\mathrm{\Gamma} \, \mathrm{d} \Gamma 
+ \int_{\mathrm{\Gamma}_\mathrm{wall}} \hat{v}_\mathrm{i} v_\mathrm{i}
+ \hat{p} \frac{\partial p}{\partial n} \mathrm{d} \Gamma  + \int_{\Omega_\mathrm{O}} j_\mathrm{\Omega} \, \mathrm{d} \Omega \nonumber \\
& + \int_{\Omega} \hat{v}_\mathrm{1}  \left[ v_\mathrm{1} \frac{\partial v_\mathrm{1}}{\partial x_\mathrm{1}} + v_\mathrm{2} \frac{\partial v_\mathrm{1}}{\partial x_\mathrm{2}} + \frac{\partial p}{\partial x_\mathrm{1}} \frac{1}{\rho}  - \nu \frac{\partial^2 v_\mathrm{1}}{\partial {x_\mathrm{2}}^2} \right] \nonumber \\
& \hspace{1cm} + \hat{v}_\mathrm{2} \left[  \frac{\partial p}{\partial x_\mathrm{2}} \frac{1}{\rho} \right]  + \hat{p} \left[ \frac{\partial v_\mathrm{1}}{\partial x_\mathrm{1}} + \frac{\partial v_\mathrm{2}}{\partial x_\mathrm{2}} \right] \mathrm{d} \Omega \; .
\end{align}
%
Again, first order optimality conditions force vanishing derivatives of the Lagrangian in all dependent directions
($\delta_{\mathrm{v}_\mathrm{i}} L \cdot \delta v_\mathrm{i} =\delta_\mathrm{v_\mathrm{1}} L \cdot \delta v_\mathrm{1} 
+ \delta_\mathrm{v_\mathrm{2}} L \cdot \delta v_\mathrm{2}  = 0 \ \forall \left(  \delta v_\mathrm{1}, \delta v_\mathrm{2} \right)$, $\delta_\mathrm{p} L \cdot \delta p = 0 \ \forall \delta p$), i.e.
\begin{align}
\delta_\mathrm{v_\mathrm{1}} L \cdot \delta v_\mathrm{1} 
+ \delta_\mathrm{v_\mathrm{2}} L \cdot \delta v_\mathrm{2} 
&= \int_{\mathrm{\Gamma}_\mathrm{wall}} \hat{v}_\mathrm{1} \delta v_\mathrm{1} + \hat{v}_\mathrm{2} \delta v_\mathrm{2}  \mathrm{d} \Gamma \nonumber \\
&\hspace{5mm} + \int_{\Gamma} \delta v_\mathrm{1}  \left[ \frac{\partial j_\mathrm{\Gamma}}{\partial v_\mathrm{1}}  + \hat{v}_\mathrm{1} \frac{\partial v_\mathrm{1}}{\partial x_\mathrm{1}} + \hat{v}_\mathrm{1} v_\mathrm{1} n_\mathrm{1} + \hat{v}_\mathrm{1} v_\mathrm{2} n_\mathrm{2} - \hat{p} n_\mathrm{2} + \nu \frac{\partial \hat{v}_\mathrm{1}}{\partial x_\mathrm{2}} n_\mathrm{2}  \right] 
- \nu \left( \hat{v}_\mathrm{1} \frac{\partial (\delta v_\mathrm{1})}{\partial x_\mathrm{2}} n_\mathrm{2} \right) \nonumber \\
&\hspace{10mm} + \delta v_\mathrm{2} \left[ \frac{\partial j_\mathrm{\Gamma}}{\partial v_\mathrm{2}} - \hat{p} n_\mathrm{2} \right] \mathrm{d} \Gamma \nonumber \\
&\hspace{5mm} + \int_{\Omega} \delta v_\mathrm{1}  \left[ \hat{v}_\mathrm{1} \frac{\partial v_\mathrm{1}}{\partial x_\mathrm{1}} - \frac{ \partial \hat{v}_\mathrm{1} v_\mathrm{1}}{\partial x_\mathrm{1}} - \frac{\partial \hat{v}_\mathrm{1} v_\mathrm{2} }{\partial x_\mathrm{2}} - \nu \frac{\partial^2 \hat{v}_\mathrm{1}}{\partial {x_\mathrm{2}}^2} - \frac{\partial \hat{p}}{\partial x_\mathrm{1}}  + \frac{\partial j_\mathrm{\Omega}}{\partial v_\mathrm{1}}\right] \nonumber \\
&\hspace{10mm} + \delta v_\mathrm{2}  \left[ \hat{v}_\mathrm{1} \frac{\partial v_\mathrm{1}}{\partial x_\mathrm{2}} - \frac{\partial \hat{p}}{\partial x_\mathrm{2}} + \frac{\partial j_\mathrm{\Omega}}{\partial v_\mathrm{2}}\right] \mathrm{d} \Omega \nonumber \\
& \overset{!}{=} 0 \ \forall \left(  \delta v_\mathrm{1}, \delta v_\mathrm{2} \right) \label{equ:adjoint_bounday_layer_mome1_second} \\
\delta_\mathrm{p} L \cdot \delta p
& = \int_{\mathrm{\Gamma}_\mathrm{wall}} \hat{p} \frac{\partial (\delta p)}{\partial n} \mathrm{d} \Gamma \nonumber \\
&\hspace{5mm} + \int_{\Gamma} \delta p \left[\frac{\partial j_\mathrm{\Gamma}}{\partial p} + \hat{v}_\mathrm{1} n_\mathrm{1} \frac{1}{\rho} +\hat{v}_\mathrm{2} n_\mathrm{2} \frac{1}{\rho}  \right] \mathrm{d} \Gamma + \int_{\Omega} \delta p  \left[ - \frac{\partial \hat{v}_\mathrm{1}}{\partial x_\mathrm{1}} \frac{1}{\rho} - \frac{\partial \hat{v}_\mathrm{2}}{\partial x_\mathrm{2}} \frac{1}{\rho} \right] \mathrm{d} \Omega \nonumber \\
& \overset{!}{=} 0 \ \forall \delta p. \label{equ:adjoint_bounday_layer_mass_second}
\end{align}
The adjoint b.-l. equations as well as their boundary conditions in Tab. (\ref{tab:bound_condi}) are obtained from Eqns. (\ref{equ:adjoint_bounday_layer_mome1_second})-(\ref{equ:adjoint_bounday_layer_mass_second}).
Assuming a sufficiently smooth boundary, a sensitivity rule w.r.t. a generalized control parameter can be computed based on remaining first order optimality criteria. Basically this control-derivative depends on the definition of the control. In the context of shape optimisation, we exemplary refer to a shape-derivative along the controlled design wall \cite{stuck2013adjoint, kuhl2019decoupling, kuhl2020adjoint} 
\begin{align}
\delta_u J 
=  \int_{\Gamma_\mathrm{D}} s (V_\mathrm{i} n _\mathrm{i}) \, \mathrm{d} \Gamma_\mathrm{O} 
= \int_{\Gamma_\mathrm{D}} s \, \mathrm{d} \Gamma_\mathrm{O} 
\qquad \mathrm{with} \qquad 
s = - \nu \frac{\partial v_\mathrm{i}}{\partial x_\mathrm{j}} \frac{\partial \hat{v_\mathrm{i}}}{\partial x_k} n_\mathrm{j} n_k \label{equ:shape_derivative},
\end{align}
in the direction of a pseudo-velocity field $V_\mathrm{i}$ (cf. \cite{kuhl2019decoupling}). In this work we confine our-self to $V_\mathrm{i} = n_\mathrm{i}$.

%
\section{Derivation of the Adjoint Blasius Equation}
\label{sec:adjoint_blasius_equation}
\begin{figure}
\centering
\begin{tikzpicture}
      \def \yStart{0.0}
      \def \yEnd{3.0}
      \def \xStart{-2.0}
      \def \xEnd{15.5}
      \def \ny{15};
      \def \xpos{-1};
      \def \dx{-0.5};

      \filldraw[pattern=north east lines, pattern color=black,draw=white] (0.0,\yStart) rectangle (\xEnd-2.0,\yStart-0.1);
      \draw [thick,dash dot] (\xStart,0.0) -- (\xEnd,0.0);
      \draw [<->] (0.0,-0.3) -- (\xEnd-2.0,-0.3);
      \draw [] (6.5,-0.7) node[right] {$L$};

      \draw[-stealth] (0.0,0.0) -- (\xEnd-1.5,0.0) node[above left] {$x_\mathrm{1}$};
      \draw[-stealth] (0.0,0.0) -- (0.0,\yEnd) node[below left] {$x_\mathrm{2}$};

      \filldraw[fill=lightgray] (\xpos+\dx,0.0) rectangle (\xpos,\yEnd) node[above] {$V_\mathrm{1}$};
      \foreach \y in {1,...,\ny}{
            \draw[-stealth] (\xpos+\dx,0.0 + \yEnd/\ny*\y) -- (\xpos,0.0+ \yEnd/\ny*\y);
      }

      \def \xpos{0.5};
      \def \ypos{0};
      \def \dx{0.5};
      \def \dy{1};
      \filldraw[fill=lightgray,name path=curve] (\xpos,\ypos)..controls(\xpos+\dx,\ypos) and(\xpos+\dx,\ypos+\dy/2) ..(\xpos+\dx,\ypos+\dy) 
             -- (\xpos+\dx,\yEnd)
             -- (\xpos,\yEnd)
             -- (\xpos,\ypos);
      \foreach \y in {1,...,\ny}{
            \path[name path=horizontal] (\xpos,\ypos + \yEnd/\ny*\y) -- + (1,0);
            \draw[-stealth,name intersections={of=curve and horizontal}] (\xpos,\ypos + \yEnd/\ny*\y) -- (intersection-1);
      }
      \def \xpos{1.5};
      \def \ypos{0};
      \def \dx{0.5};
      \def \dy{1.414};
      \filldraw[fill=lightgray,name path=curve] (\xpos,\ypos)..controls(\xpos+\dx,\ypos) and(\xpos+\dx,\ypos+\dy/2) ..(\xpos+\dx,\ypos+\dy) 
             -- (\xpos+\dx,\yEnd)
             -- (\xpos,\yEnd)
             -- (\xpos,\ypos);
      \foreach \y in {1,...,\ny}{
            \path[name path=horizontal] (\xpos,\ypos + \yEnd/\ny*\y) -- + (1,0);
            \draw[-stealth,name intersections={of=curve and horizontal}] (\xpos,\ypos + \yEnd/\ny*\y) -- (intersection-1);
      }
      \def \xpos{2.5};
      \def \ypos{0};
      \def \dx{0.5};
      \def \dy{1.732};
      \filldraw[fill=lightgray,name path=curve] (\xpos,\ypos)..controls(\xpos+\dx,\ypos) and(\xpos+\dx,\ypos+\dy/2) ..(\xpos+\dx,\ypos+\dy) 
             -- (\xpos+\dx,\yEnd)
             -- (\xpos,\yEnd)
             -- (\xpos,\ypos);
      \foreach \y in {1,...,\ny}{
            \path[name path=horizontal] (\xpos,\ypos + \yEnd/\ny*\y) -- + (1,0);
            \draw[-stealth,name intersections={of=curve and horizontal}] (\xpos,\ypos + \yEnd/\ny*\y) -- (intersection-1);
      }
      \def \xpos{3.5};
      \def \ypos{0};
      \def \dx{0.5};
      \def \dy{2};
      \filldraw[fill=lightgray,name path=curve] (\xpos,\ypos)..controls(\xpos+\dx,\ypos) and(\xpos+\dx,\ypos+\dy/2) ..(\xpos+\dx,\ypos+\dy) 
             -- (\xpos+\dx,\yEnd)
             -- (\xpos,\yEnd)
             -- (\xpos,\ypos);
      \foreach \y in {1,...,\ny}{
            \path[name path=horizontal] (\xpos,\ypos + \yEnd/\ny*\y) -- + (1,0);
            \draw[-stealth,name intersections={of=curve and horizontal}] (\xpos,\ypos + \yEnd/\ny*\y) -- (intersection-1);
      }
      \def \xpos{4.5};
      \def \ypos{0};
      \def \dx{0.5};
      \def \dy{2.236};
      \filldraw[fill=lightgray,name path=curve] (\xpos,\ypos)..controls(\xpos+\dx,\ypos) and(\xpos+\dx,\ypos+\dy/2) ..(\xpos+\dx,\ypos+\dy) 
             -- (\xpos+\dx,\yEnd)
             -- (\xpos,\yEnd)
             -- (\xpos,\ypos);
      \foreach \y in {1,...,\ny}{
            \path[name path=horizontal] (\xpos,\ypos + \yEnd/\ny*\y) -- + (1,0);
            \draw[-stealth,name intersections={of=curve and horizontal}] (\xpos,\ypos + \yEnd/\ny*\y) -- (intersection-1);
      }
      \draw[scale=1.0,domain=0:2.3,smooth,variable=\x,black]  plot ({\x*\x},{\x}) node[right] {$\sqrt{x_\mathrm{1}}$};

      \def \xpos{15.0};
      \def \dx{-0.5};
      \filldraw[fill=lightgray] (\xpos+\dx,0.0) rectangle (\xpos,\yEnd) node[above] {$\hat{V}_\mathrm{1}$};
      \foreach \y in {1,...,\ny}{
            \draw[-stealth] (\xpos+\dx,0.0 + \yEnd/\ny*\y) -- (\xpos,0.0+ \yEnd/\ny*\y);
      }

      \def \xpos{8.5};
      \def \ypos{0};
      \def \dx{0.5};
      \def \dy{2.236};
      \filldraw[fill=lightgray,name path=curve] (\xpos,\ypos)..controls(\xpos+\dx,\ypos) and(\xpos+\dx,\ypos+\dy/2) ..(\xpos+\dx,\ypos+\dy) 
             -- (\xpos+\dx,\yEnd)
             -- (\xpos,\yEnd)
             -- (\xpos,\ypos);
      \foreach \y in {1,...,\ny}{
            \path[name path=horizontal] (\xpos,\ypos + \yEnd/\ny*\y) -- + (1,0);
            \draw[-stealth,name intersections={of=curve and horizontal}] (\xpos,\ypos + \yEnd/\ny*\y) -- (intersection-1);
      }
      \def \xpos{9.5};
      \def \ypos{0};
      \def \dx{0.5};
      \def \dy{2};
      \filldraw[fill=lightgray,name path=curve] (\xpos,\ypos)..controls(\xpos+\dx,\ypos) and(\xpos+\dx,\ypos+\dy/2) ..(\xpos+\dx,\ypos+\dy) 
             -- (\xpos+\dx,\yEnd)
             -- (\xpos,\yEnd)
             -- (\xpos,\ypos);
      \foreach \y in {1,...,\ny}{
            \path[name path=horizontal] (\xpos,\ypos + \yEnd/\ny*\y) -- + (1,0);
            \draw[-stealth,name intersections={of=curve and horizontal}] (\xpos,\ypos + \yEnd/\ny*\y) -- (intersection-1);
      }
      \def \xpos{10.5};
      \def \ypos{0};
      \def \dx{0.5};
      \def \dy{1.732};
      \filldraw[fill=lightgray,name path=curve] (\xpos,\ypos)..controls(\xpos+\dx,\ypos) and(\xpos+\dx,\ypos+\dy/2) ..(\xpos+\dx,\ypos+\dy) 
             -- (\xpos+\dx,\yEnd)
             -- (\xpos,\yEnd)
             -- (\xpos,\ypos);
      \foreach \y in {1,...,\ny}{
            \path[name path=horizontal] (\xpos,\ypos + \yEnd/\ny*\y) -- + (1,0);
            \draw[-stealth,name intersections={of=curve and horizontal}] (\xpos,\ypos + \yEnd/\ny*\y) -- (intersection-1);
      }
      \def \xpos{11.5};
      \def \ypos{0};
      \def \dx{0.5};
      \def \dy{1.414};
      \filldraw[fill=lightgray,name path=curve] (\xpos,\ypos)..controls(\xpos+\dx,\ypos) and(\xpos+\dx,\ypos+\dy/2) ..(\xpos+\dx,\ypos+\dy) 
             -- (\xpos+\dx,\yEnd)
             -- (\xpos,\yEnd)
             -- (\xpos,\ypos);
      \foreach \y in {1,...,\ny}{
            \path[name path=horizontal] (\xpos,\ypos + \yEnd/\ny*\y) -- + (1,0);
            \draw[-stealth,name intersections={of=curve and horizontal}] (\xpos,\ypos + \yEnd/\ny*\y) -- (intersection-1);
      }
      \def \xpos{12.5};
      \def \ypos{0};
      \def \dx{0.5};
      \def \dy{1};
      \filldraw[fill=lightgray,name path=curve] (\xpos,\ypos)..controls(\xpos+\dx,\ypos) and(\xpos+\dx,\ypos+\dy/2) ..(\xpos+\dx,\ypos+\dy) 
             -- (\xpos+\dx,\yEnd)
             -- (\xpos,\yEnd)
             -- (\xpos,\ypos);
      \foreach \y in {1,...,\ny}{
            \path[name path=horizontal] (\xpos,\ypos + \yEnd/\ny*\y) -- + (1,0);
            \draw[-stealth,name intersections={of=curve and horizontal}] (\xpos,\ypos + \yEnd/\ny*\y) -- (intersection-1);
      }

      \draw[scale=1.0,domain=0:2.3,smooth,variable=\x,black]  plot ({13.5-\x*\x},{\x}) node[left] {$\sqrt{L - x_\mathrm{1}}$};
\end{tikzpicture}
\caption{Illustration of the (forward) primal $V_1$ and (backward) adjoint $\hat V_1$ flow over a flat finite plate in the $x_\mathrm{1}$-$x_\mathrm{2}$ plane.}
\label{fig:plate_flow}
\end{figure}
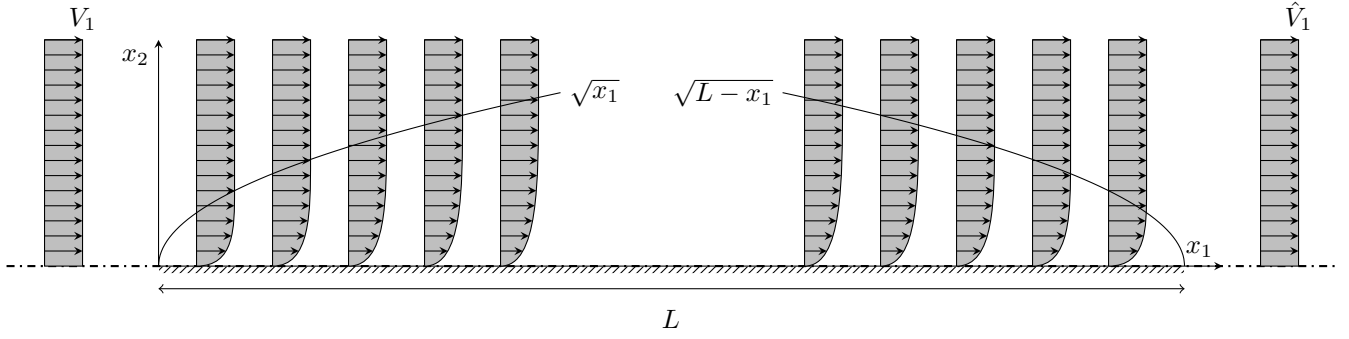

Prior to addressing the adjoint Blasius solution, it is instructive to repeat 
the fundamentals of the primal Blasius solution in brief. Further details can be found in textbooks, e.g. \cite{schlichting2006grenzschicht}.
%
We consider the flat plate b.-l. flow where the extent in streamwise $x_\mathrm{1}$ is much larger than in $x_\mathrm{2}$ direction normal to the plate, cf. Fig. \ref{fig:plate_flow}. The plate length $L$ is expected to tend to infinity, i.e. $L \rightarrow \infty$, and we assume homogeneous steady inflow $V_1$.
Generalized solutions for Eqns. (\ref{equ:primal_bounday_layer_mome1_first})-(\ref{equ:primal_bounday_layer_mass1_first}) try to downgrade the PDEs to ODEs. The latter is achieved based on a coordinate transformation, which in turn needs suited similarity coordinates.
For both, the primal and the adjoint system, we conceptually split the derivation in two parts: (a) we derive a suitable similarity variable that (b) should simplify the corresponding equations.

\paragraph{The primal boundary layer} 
 flow is anticipated to be a function of the plate normal and tangential coordinate, viz. $v_\mathrm{1} / V_\mathrm{1} = g ( \eta)$ with $\eta = x_\mathrm{2} / \delta(x_\mathrm{1})$, where $g$ and $\delta$ represent a (so far) unknown function as well as a measure for the b.-l. thickness respectively. Thus, we can directly compute its spatial derivatives, e.g. $\partial v_\mathrm{1} / \partial x_\mathrm{1} = - V_\mathrm{1} g^\prime ( x_\mathrm{2} / \delta^2 ) ( \partial \delta / \partial x_\mathrm{1} )$ with $g^\prime = \mathrm{d} g / \mathrm{d} \eta$.
An integration of Eqn. (\ref{equ:primal_bounday_layer_mome1_first}) along the wall-normal coordinate from the wall to the b.-l. edge reads 
\begin{align}
2 \int_0^\delta v_\mathrm{1}  \frac{\partial v_\mathrm{1} }{\partial x_\mathrm{1} } \mathrm{d} x_\mathrm{2}  
+ \int_0^\delta v_\mathrm{2}  \frac{\partial v_\mathrm{1} }{\partial x_\mathrm{2} } \mathrm{d} x_\mathrm{2} 
- \int_0^\delta \nu \frac{\partial^2 v_\mathrm{1} }{\partial x_\mathrm{2} ^2} \mathrm{d} x_\mathrm{2}   = 0.
\end{align}
Applying the continuity equation ($\partial v_\mathrm{2} / \partial x_\mathrm{2} = -\partial v_\mathrm{1} / \partial x_\mathrm{1}$) together with the new definition of the plate tangential velocity allows the elimination of the plate-normal velocity, i.e. 
\begin{align}
-2 V_\mathrm{1}^2 \int_0^\delta  g g^\prime \frac{x_\mathrm{2}}{\delta^2} \frac{\partial \delta}{\partial x_\mathrm{1}} \mathrm{d} x_\mathrm{2} 
+ V_\mathrm{1}^2  \int_0^\delta  g^\prime \frac{x_\mathrm{2} }{\delta^2} \frac{\partial \delta}{\partial x_\mathrm{1}} \mathrm{d} x_\mathrm{2} 
- \nu V_\mathrm{1} g^\prime \frac{1}{\delta} \bigg|_0^\delta  = 0.
\end{align}
Substituting $\eta = x_\mathrm{2} / \delta$ ($\mathrm{d} x_\mathrm{2} = \delta \mathrm{d} \eta = \mathrm{d} (x_\mathrm{2} / \delta)$) offers an ODE for $\delta$, viz.
\begin{align}
\left[B - 2 A \right] \frac{\partial \delta}{\partial x_\mathrm{1}} \delta  = \frac{\nu}{V_\mathrm{1}} C \label{equ:primal_delta_ode},
\end{align}
where all integrals are condensed to the parameters $A$ and $B$, hence  
\begin{align}
A = \int_0^1  g g^\prime \eta \, \mathrm{d} \eta \hspace{1cm} B  = \int_0^1  g^\prime \eta \,  \mathrm{d} \eta \hspace{1cm} C = g^\prime(1) - g^\prime(0).
\end{align}
Eqn. (\ref{equ:primal_delta_ode}) can be solved w.r.t. $\delta$, which yields 
\begin{align}
\delta = \sqrt{\frac{2 C}{b \left[ B - 2 A \right] }} \sqrt{ \frac{\nu \left[ a + b x_\mathrm{1} \right]}{V_\mathrm{1}}} 
\hspace{1cm} \Rightarrow \hspace{1cm} 
\delta \propto \sqrt{ \frac{\nu \left[a + b x_\mathrm{1} \right]}{V_\mathrm{1}} } \label{equ:primal_delta} \; . 
\end{align}
The last expression supports an estimation of the boundary-layer thickness and offers a suitable choice for the similarity variable $\eta = x_\mathrm{2} \sqrt{V_\mathrm{1} / [ \nu (a + b x_\mathrm{1})]}$.
It should be pointed out that classical b.-l. thickness measures employ $a=0$ and $b=1$ which is basically due to the choice for the origin of the coordinate system. However, motivated by the subsequent adjoint analysis, we continue with the more general expression.

Since a suitable similarity variable is found, we seek for a velocity field that satisfies Eqns. (\ref{equ:primal_bounday_layer_mome1_first})-(\ref{equ:primal_bounday_layer_mass1_first}). Introducing a stream function $\psi$ that inherently satisfies the continuity expression (\ref{equ:primal_bounday_layer_mass1_first}), e.g. $v_\mathrm{1} = \partial \psi / \partial x_\mathrm{2}$, $v_\mathrm{2} = -\partial \psi / \partial x_\mathrm{1}$, offers access to the plate tangential velocity, viz.
\begin{align}
\psi 
&= \int_0^{x_\mathrm{2}} v_\mathrm{1} \mathrm{d} x_\mathrm{2} 
= \int_{\eta(0)}^{\eta(x_\mathrm{2})} g (\eta) V_\mathrm{1} \delta \mathrm{d} \eta
= \sqrt{\nu \left[a + b x_\mathrm{1} \right] V_\mathrm{1}} \underbrace{\int_0^\eta g  \mathrm{d} \eta}_{f(\eta)} \label{equ:primal_stream_function}.
\end{align}
All primal b.-l. terms can be expressed in terms of $f$ and $\eta$, e.g. $\partial v_\mathrm{1}^2 / \partial x_\mathrm{2}^2 = V_\mathrm{1}^2 / [\nu (a + b x_\mathrm{1})] f^{\prime \prime \prime}$, as summarized in App. \ref{app:similarity_relations}. Assuming a homogeneous pressure field, the substitutions of all terms of Eqn. (\ref{equ:primal_bounday_layer_mome1_first}) yields the well known Blasius equation 
\begin{align}
R_\mathrm{1}^\mathrm{BL} \qquad &\rightarrow \qquad  - 2 f^{\prime \prime \prime} -  b f  f^{\prime \prime} = 0  \label{equ:primal_blasius_equation}.
\end{align}

\paragraph{The adjoint boundary layer}
quantities are convected by the primal flow and we anticipate a similar behavior of the adjoint field and a scalar field, e.g. a temperature field, although the latter usually does not have to fulfill a continuity equation. The interpretation of the adjoint flow is non-intuitive due to the influence of the objective functional, e.g. volume based objectives possibly introduce a non-divergence free adjoint velocity field. Therefore we confine the present investigations to boundary based objectives.
In line with the primal flow, we start with the similarity variable and assume the adjoint mean flow to depend on the plate normal and tangential coordinate, viz. $\hat{v}_\mathrm{1} / \hat{V}_\mathrm{1} = \hat{g} ( \hat{\eta})$ with $\hat{\eta} = x_\mathrm{2} / \hat{\delta}(x_\mathrm{1})$ where $\hat{g}$ and $\hat{\delta}$ represent the adjoint complement of $g$ and $\delta$. Again, we can directly compute all required spatial derivatives, e.g. $\partial \hat{v}_\mathrm{1} / \partial x_\mathrm{1} = - \hat{V}_\mathrm{1} \hat{g}^\prime ( x_\mathrm{2} / \hat{\delta}^2 ) ( \partial \hat{\delta} / \partial x_\mathrm{1} )$ with $\hat{g}^\prime = \mathrm{d} \hat{g} / \mathrm{d} \hat{\eta}$.
An integration of (\ref{equ:adjoint_bounday_layer_mome1_first}) between the wall and the adjoint b.-l. edge while assuming adjoint ZPG along the wall normal coordinate reads
\begin{align}
-\int_0^{\hat{\delta}} v_\mathrm{1}  \frac{\partial \hat{v}_\mathrm{1} }{\partial x_\mathrm{1} } \mathrm{d} x_\mathrm{2}  
- \int_0^{\hat{\delta}} v_\mathrm{2}  \frac{\partial \hat{v}_\mathrm{1} }{\partial x_\mathrm{2} } \mathrm{d} x_\mathrm{2} 
- \int_0^{\hat{\delta}} \nu \frac{\partial^2 \hat{v}_\mathrm{1} }{\partial x_\mathrm{2} ^2} \mathrm{d} x_\mathrm{2} 
= - \int_0^{\hat{\delta}} \hat{v}_\mathrm{1}  \frac{\partial v_\mathrm{1} }{\partial x_\mathrm{1} } \mathrm{d} x_\mathrm{2} \; .
\end{align}
Here the additional term on the right hand side corresponds to the ATC term originating from the non-linear convection.
Applying the primal continuity equation ($\partial v_\mathrm{2} / \partial x_\mathrm{2} = -\partial v_\mathrm{1} / \partial x_\mathrm{1}$) interestingly cancels the ATC term, viz.
\begin{align}
- \int_0^{\hat{\delta}} v_\mathrm{1}  \frac{\partial \hat{v}_\mathrm{1} }{\partial x_\mathrm{1} } \mathrm{d} x_\mathrm{2}  
+ \hat{V}_\mathrm{1} \int_0^{\hat{\delta}} \frac{\partial v_\mathrm{1}}{\partial x_\mathrm{1}} \mathrm{d} x_\mathrm{2} 
- \int_0^{\hat{\delta}} \hat{v}_\mathrm{1} \frac{\partial v_\mathrm{1}}{\partial x_\mathrm{1}} \mathrm{d} x_\mathrm{2} 
- \nu \frac{\partial \hat{v}_\mathrm{1}}{\partial x_\mathrm{2}} \bigg|_0^{\hat{\delta}} 
= - \int_0^{\hat{\delta}} \hat{v}_\mathrm{1}  \frac{\partial v_\mathrm{1}}{\partial x_\mathrm{1}} \mathrm{d} x_\mathrm{2} 
\; . 
\end{align}
Combining the primal tangential velocity ($v_\mathrm{1} / V_\mathrm{1} = g(\eta)$) and  
 its (anticipated) adjoint complement yields
\begin{align}
\frac{\partial \hat{\delta}}{\partial x_\mathrm{1}} \int_0^1 g  \hat{g}^\prime \hat{\eta} \mathrm{d} \hat{\eta}
- \frac{\partial \delta}{\partial x_\mathrm{1}} \int_0^{\delta / \hat{\delta}} g^\prime  \eta  \mathrm{d} \eta
- \frac{\nu}{V_\mathrm{1}} \hat{g}^\prime \frac{1}{\hat{\delta}} \bigg|_0^1= 0.
\end{align}
Substituting $\hat{\eta} = x_\mathrm{2} / \hat{\delta}$ ($\mathrm{d} x_\mathrm{2} = \hat{\delta} \mathrm{d} \hat{\eta} = \hat{\delta} \mathrm{d} (x_\mathrm{2} / \hat{\delta})$) offers an ODE for $\hat{\delta}$ that inheres the primal b.-l. thickness measure
\begin{align}
\left[ \hat{A} \frac{\partial \hat{\delta} }{\partial x_\mathrm{1}} - \hat{B} \frac{\partial \delta}{\partial x_\mathrm{1}} \right] \hat{\delta}  =  \frac{\nu}{V_\mathrm{1}} \hat{C}  \label{equ:adjoint_delta_ode}.
\end{align}
Again, all integrals are condensed into coefficients, viz. 
\begin{align}
\hat{A} = \int_0^1 g  \hat{g}^\prime \hat{\eta} \ \mathrm{d} \hat{\eta} \hspace{1cm} \hat{B} = \int_0^{\delta / \hat{\delta}} g^\prime \eta \  \mathrm{d} \eta \hspace{1cm} \hat{C} &= \hat{g}^\prime(1) - \hat{g}^\prime(0).
\end{align}
Thanks to the available primal $\delta$ we can reformulate Eqn. (\ref{equ:adjoint_delta_ode}), viz.
\begin{align}
C_\mathrm{1} \frac{\partial \hat{\delta} }{\partial x_\mathrm{1}} \hat{\delta}  - C_\mathrm{2} \frac{\hat{\delta}  }{\sqrt{a + b x_\mathrm{1}}}  =  C_\mathrm{3} \label{equ:adjoint_delta_ode_simplified}
\end{align}
where $C_\mathrm{1} = \hat{A}$, $C_\mathrm{2} = \hat{B} (b/2)  \sqrt{2 C / (b [A-2B]) } \sqrt{\nu / V_\mathrm{1}} $ and $C_\mathrm{3} = \hat{C} \nu / V_\mathrm{1}$ are introduced to keep the notation compact. Thus, we can solve Eqn. (\ref{equ:adjoint_delta_ode_simplified}) w.r.t. $\hat{\delta}$
\begin{align}
\hat{\delta} = -\frac{2 C_\mathrm{3}}{C_\mathrm{2} + \sqrt{C_\mathrm{2}^2 + 2 b C_\mathrm{1} C_\mathrm{3}}} \sqrt{a + b x_\mathrm{1}}
\end{align}
and quantify an adjoint b.-l. thickness by the expression 
\begin{align}
\hat{\delta} &= - \frac{C_\mathrm{3}}{C_\mathrm{2}} \frac{b \hat{B}}{C_\mathrm{2} + \sqrt{C_\mathrm{2}^2 + 2 b C_\mathrm{1} C_\mathrm{3}}} \underbrace{ \sqrt{\frac{2C}{b \left[B-2 A \right]}} \sqrt{ \frac{\nu (a + bx)}{V_\mathrm{1}}} }_{\delta} 
\hspace{1cm} \Rightarrow \hspace{1cm} 
\hat{\delta} \propto \delta \propto \sqrt{ \frac{\nu (a+bx)}{V_\mathrm{1}} } \label{equ:adjoint_delta},
\end{align}
which is somehow proportional to the primal b.-l. thickness.
Finally we can define an adjoint similarity variable that equals the primal one, viz.
\begin{align}
\hat{\eta} &= \frac{x_2}{\hat{\delta}} \propto x_2 \sqrt{\frac{V_\mathrm{1}}{\nu (a+b x_1)}} = \frac{x_2}{\delta} = \eta \label{equ:adjoint_similarity}.
\end{align}
A challenge in the interpretation of the adjoint results follows from the inverted convection characteristics of the adjoint flow field.
To describe the primal b.-l. velocity, the origin of the coordinate system is typically positioned at the plate origin (a=0,b=1). However, a reasonable measurement based on the plate end is obtained for the adjoint system (a=L,b=-1). 
The latter motivates the negative sign in Eqn. (\ref{equ:adjoint_delta}) and indicates a formal issue for the numerical verification/validation of the approach: Physically the plate is assumed to extent infinitely in streamwise  direction. For an adjoint approach, the latter is numerically uncomfortable.

Analogous to the primal derivation, we seek now for an adjoint velocity field that satisfies Eqns. (\ref{equ:adjoint_bounday_layer_mome1_first})-(\ref{equ:adjoint_bounday_layer_mass_first}) where the primal flow field is already known.
Again, we have to satisfy a continuity equation and therefore define an adjoint stream function $\hat{\psi}$ that inherently complies with (\ref{equ:adjoint_bounday_layer_mass_first}), e.g. $\hat{v}_\mathrm{1} = \partial \hat{\psi} / \partial x_\mathrm{2}$, $\hat{v}_\mathrm{2} = -\partial \hat{\psi} / \partial x_\mathrm{1}$ and offers access to the wall tangential adjoint velocity, viz.
\begin{align}
\hat{\psi} 
= \int_0^{x_\mathrm{2}} \hat{v}_\mathrm{1} \mathrm{d} x_\mathrm{2}
= \int_{\hat{\eta}(0)}^{\hat{\eta}(x_\mathrm{2})} \hat{g} (\hat{\eta}) \hat{V}_\mathrm{1} \hat{\delta} \mathrm{d} \hat{\eta}
= \sqrt{ \frac{\nu \left[a + b x_\mathrm{1} \right] \hat{V}_\mathrm{1}^2}{V_\mathrm{1}}} \underbrace{\int_0^{\hat{\eta}} \hat{g}  \mathrm{d} \hat{\eta}}_{\hat{f}(\hat{\eta}) }.
\end{align}
All adjoint b.-l. expression can be computed from $\hat f$ and $\hat \eta$, e.g. $\partial \hat{v}_\mathrm{1}^2 / \partial x_\mathrm{2}^2 = \hat{V}_\mathrm{1} V_\mathrm{1} / [\nu (a + b x_\mathrm{1})] \hat{f}^{\prime \prime \prime}$. Details of the  similarity transformations are provided in App. \ref{app:similarity_relations}. The streamwise adjoint b.-l. equation (\ref{equ:adjoint_bounday_layer_mome1_first}) reduces to
\begin{align}
 b \left[ \hat{\eta} - \eta \right] f^\prime  \hat{f}^{\prime \prime}
+ b f  \hat{f}^{\prime \prime}  
- 2 \hat{f}^{\prime \prime \prime} 
= b f^{\prime \prime} \hat{f}^\prime \eta \; . 
\end{align}
In combination with (\ref{equ:adjoint_similarity}) we achieve the adjoint complement to the Blasius equation
\begin{alignat}{3}
\hat{R}_\mathrm{1}^\mathrm{BL} \qquad &\rightarrow \qquad
- 2 \hat{f}^{\prime \prime \prime} 
+ &&b \hat{f}^{\prime \prime}  f 
&&= b \hat{f}^\prime f^{\prime \prime} \eta \label{equ:tangential_adjoint_blasius_equation} \; , \\
\hat{R}_\mathrm{2}^\mathrm{BL} \qquad &\rightarrow \qquad
&&\hat{f}^\prime f^{\prime \prime} &&= 0  \label{equ:normal_adjoint_blasius_equation} \; . 
\end{alignat}
 The first [second] equation corresponds to the generalized tangential [normal] adjoint b.-l. equation. 
Interestingly, the normal adjoint momentum balance (\ref{equ:normal_adjoint_blasius_equation}) cancels the ATC term in the corresponding tangential direction (\ref{equ:tangential_adjoint_blasius_equation}).
The latter is similar to their primal counterpart with an inverted sign in front of the non-linearity originating from the inverted convection characteristics.

To pursue the \textit{simplify-and-derive} strategy, one typically first inserts the primal simplification. As a result, however, the streamwise adjoint Blasius equation (\ref{equ:tangential_adjoint_blasius_equation}) cannot be retrieved directly from a variation of (\ref{equ:primal_blasius_equation}), due to the ($x_\mathrm{1}$-nonlinear) $\eta$-based coordinate transformation inherent to the initial simplification step.
A way out avoids the similarity transformation and declares the $x_\mathrm{2}$-derivative of the adjoint stream function $\hat{\psi}_{x_\mathrm{2}} = \partial \hat{\psi} / \partial x_\mathrm{2}$ as Lagrange multiplier. The latter represents an educated guess, which follows from the previous discussions of the paper, viz.
\begin{align}
L = ... 
+ \int_\mathrm{x_\mathrm{1}} \int_\mathrm{x_\mathrm{2}} 
\hat{v}_\mathrm{1} R_\mathrm{1}^\mathrm{BL} \,
\mathrm{d} x_\mathrm{2} \, \mathrm{d} x_\mathrm{1} 
\qquad \rightarrow \qquad 
L = ...
+ \int_\mathrm{x_\mathrm{1}} \int_\mathrm{x_\mathrm{2}} 
\hat{\psi}_{x_\mathrm{2}} \left[ 
\psi_{x_\mathrm{2}} \psi_{x_\mathrm{2}, x_\mathrm{1}} 
-\psi_{x_\mathrm{1}} \psi_{x_\mathrm{2},x_\mathrm{2}} 
+ \nu \, \psi_{x_\mathrm{2},x_\mathrm{2},x_\mathrm{2}} \right]
\mathrm{d} x_\mathrm{2} \, \mathrm{d} x_\mathrm{1}.
\label{eq:wayout}
\end{align}
Using first order optimality conditions, the adjoint equations can be derived from the stream function based formulation (\ref{eq:wayout}) using integration by parts in Euclidean (Cartesian) space 
\begin{align}
\delta_{\psi_{(x_\mathrm{2})}} L \cdot \delta (\psi_{x_\mathrm{2}})  &= ... 
+ \int_\mathrm{x_\mathrm{1}} \int_\mathrm{x_\mathrm{2}} 
\delta \left( \psi_{x_\mathrm{2}} \right) \left[ 
\hat{\psi}_{x_\mathrm{2}}  \psi_{x_\mathrm{2}, x_\mathrm{1}} 
- \left(\hat{\psi}_{x_\mathrm{2}} \psi_{x_\mathrm{2}} \right)_{x_\mathrm{1}} 
- \left(\hat{\psi}_{x_\mathrm{2}} \psi_{x_\mathrm{1}} \right)_{x_\mathrm{2}} 
+ \nu \, \hat{\psi}_{x_\mathrm{2},x_\mathrm{2},x_\mathrm{2}} \right]
\mathrm{d} x_\mathrm{2} \, \mathrm{d} x_\mathrm{1} \; , \\
\delta_{\psi_{(x_\mathrm{1})}} L \cdot \delta ( \psi_{x_\mathrm{1}} )  & = ... 
-\int_\mathrm{x_\mathrm{1}} \int_\mathrm{x_\mathrm{2}}
\delta \left( \psi_{x_\mathrm{1}} \right) \left[ 
\hat{\psi}_{x_\mathrm{2}} \psi_{x_\mathrm{2},x_\mathrm{2}} \right]
\mathrm{d} x_\mathrm{2} \, \mathrm{d} x_\mathrm{1}b \; . 
\end{align}
We subsequently impose a similarity transformation based on available primal and adjoint similarity relations, viz.
\begin{alignat}{2}
\delta_{f^\prime} L \cdot \delta (f^\prime)  &= ... +
\int_\mathrm{x_\mathrm{1}} \int_\eta 
\delta (f^\prime) \left[  - b \hat{f}^\prime f^{\prime \prime} \eta + b \hat{f}^{\prime \prime} f -2 \hat{f}^{\prime \prime \prime} \right]
\mathrm{d} \eta \sqrt{\frac{V_\mathrm{1}^3 \hat{V}_\mathrm{1}^2 \nu }{a + b x_\mathrm{1}}} \mathrm{d} x_\mathrm{1}
\qquad &&\overset{!}{=} 0 \qquad \forall \, \delta ( f^\prime) \label{equ:tangential_adjoint_blasius_equation_2} \\
\delta_{f^\prime \eta - f } L \cdot \delta (f^\prime \eta - f )  &= ... +
\int_\mathrm{x_\mathrm{1}} \int_\eta 
\delta (f^\prime \eta - f ) \left[ \hat{f}^\prime f^{\prime \prime} \right]
\mathrm{d} \eta \sqrt{\frac{V_\mathrm{1}^3 \hat{V}_\mathrm{1}^2 \nu }{a + b x_\mathrm{1}}} \mathrm{d} x_\mathrm{1}
\qquad &&\overset{!}{=} 0 \qquad \forall \, \delta (f^\prime \eta - f ).
\end{alignat}
The expressions in the square brackets correspond to the plate tangential and normal adjoint Blasius equations  (\ref{equ:tangential_adjoint_blasius_equation})-(\ref{equ:normal_adjoint_blasius_equation}).
The first terms in the respective brackets contain the first derivative $\hat{f}^\prime$ and originate from the variation of the convection ($(\delta v_\mathrm{k}) \, \partial v_\mathrm{1} / \partial x_\mathrm{k}$). These terms result in the ATC contribution to the adjoint Blasius equations. The term that inheres the second derivative $\hat{f}^{\prime \prime}$ follows from the perturbed convected primal momentum ($v_\mathrm{k} \, \partial (\delta v_\mathrm{1}) / \partial x_\mathrm{k}$) and switches sign due to integration by parts. The diffusion term refers to the third term in (\ref{equ:tangential_adjoint_blasius_equation_2}) and enters the equation analogous to the primal counterpart with a third derivative. 
Its self-adjoint character is underlined by its consistent sign in the adjoint and primal equations. 

%
\section{Numerical Solution of the Adjoint Blasius Equation}
\label{sec:numerical_blasius_approximation}
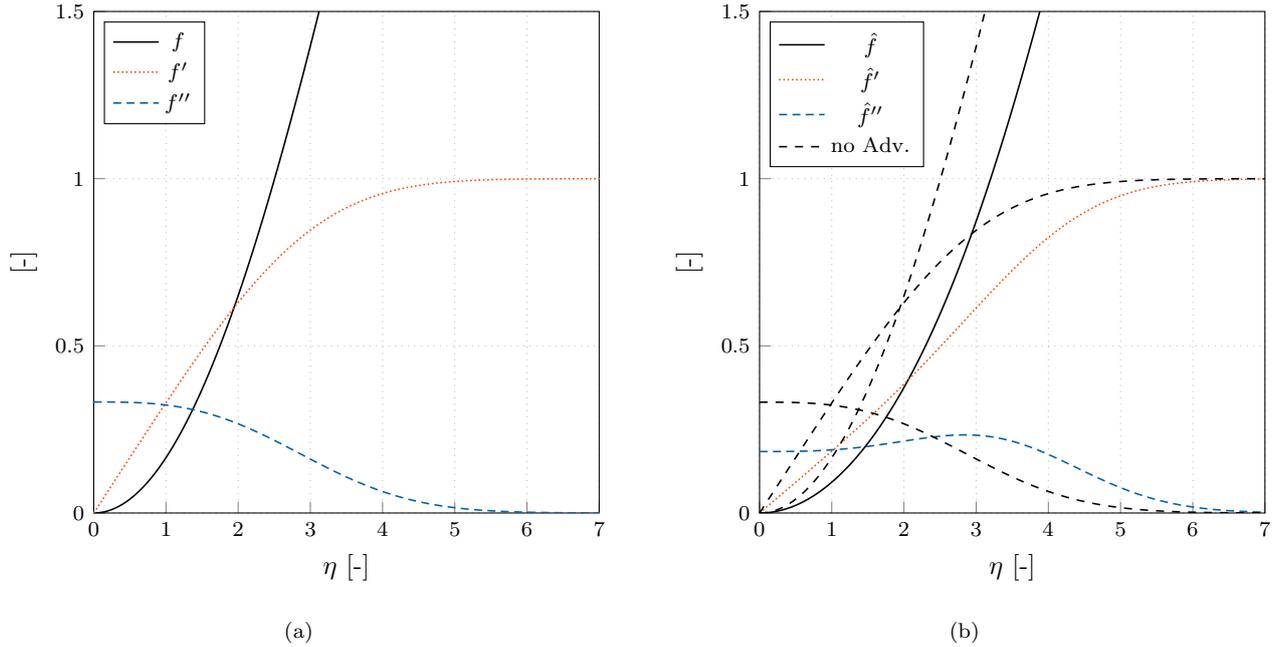
\begin{figure}
\centering
\subfigure[]{
\smallPicture
\begin{tikzpicture}
\begin{axis}[
 xlabel style={text width=0.25\textwidth,align=center},
 ylabel style={text width=0.25\textwidth,align=center},
 xlabel={$\eta$ [-]},
 ylabel={[-]},
 legend style={at={(0.02,0.98)},anchor=north west},
 xmin=0,xmax=7,
 ymin=0.0,ymax=1.5,
 xtick={0,1,2,3,4,5,6,7},
]
\addplot [line1] table[x expr={\thisrowno{0}},y expr={\thisrowno{1}}]{data/Blasius_Shooting_Primal_Results.dat};
\addplot [line2] table[x expr={\thisrowno{0}},y expr={\thisrowno{2}}]{data/Blasius_Shooting_Primal_Results.dat};
\addplot [line3] table[x expr={\thisrowno{0}},y expr={\thisrowno{3}}]{data/Blasius_Shooting_Primal_Results.dat};

\addlegendentry{$f$};
\addlegendentry{$f^{\prime}$};
\addlegendentry{$f^{\prime \prime}$};

\end{axis}
\end{tikzpicture}
}
\subfigure[]{
\smallPicture
\begin{tikzpicture}
\begin{axis}[
 xlabel style={text width=0.25\textwidth,align=center},
 ylabel style={text width=0.25\textwidth,align=center},
 xlabel={$\eta$ [-]},
 ylabel={[-]},
 legend style={at={(0.02,0.98)},anchor=north west},
 xmin=0,xmax=7,
 ymin=0.0,ymax=1.5,
 xtick={0,1,2,3,4,5,6,7},
]
\addplot [line1] table[x expr={\thisrowno{0}},y expr={\thisrowno{1}}]{data/Blasius_Shooting_Adjoint_Results_External.dat};
\addplot [line2] table[x expr={\thisrowno{0}},y expr={\thisrowno{2}}]{data/Blasius_Shooting_Adjoint_Results_External.dat};
\addplot [line3] table[x expr={\thisrowno{0}},y expr={\thisrowno{3}}]{data/Blasius_Shooting_Adjoint_Results_External.dat};
\addplot [line1, dashed] table[x expr={\thisrowno{0}},y expr={\thisrowno{1}}]{data/Blasius_Shooting_Adjoint_Results_Without_Advection_External.dat};
\addplot [line1, dashed] table[x expr={\thisrowno{0}},y expr={\thisrowno{2}}]{data/Blasius_Shooting_Adjoint_Results_Without_Advection_External.dat};
\addplot [line1, dashed] table[x expr={\thisrowno{0}},y expr={\thisrowno{3}}]{data/Blasius_Shooting_Adjoint_Results_Without_Advection_External.dat};

\addlegendentry{$\hat{f} $};
\addlegendentry{$\hat{f}^{\prime}$};
\addlegendentry{$\hat{f}^{\prime \prime}$};
\addlegendentry{no Adv.};

\end{axis}
\end{tikzpicture}
}
\caption{Results of a shooting method for (a) the primal and (b) the adjoint Blasius equation. Dashed lines indicate adjoint results without considering the adjoint transpose convection term.}
\label{fig:blasius_shooting_results}
\end{figure}

The primal/adjoint Blasius equations are numerically approximated based on a shooting method. An exemplary \copyright{Matlab} code is available at \cite{matlabblasius}. The primal procedure tries to hit the boundary value $f^\prime_\infty = 1$ for $\eta \to \infty$ with prescribed wall values $f_0 = 0$ and $f^{\prime}_0 = 0$.
The boundary value problem is controlled by $f^{\prime \prime}_0$ at the wall and iterated to convergence until the value $f^\prime_\infty $ falls below a numerical limit of $(f^\prime_\infty - 1) < 10^{-08}$.
The result is depicted by the left graph of Fig. \ref{fig:blasius_shooting_results}, where $a=0$ and $b=1$ were assumed. The displayed numerical results perfectly match with available data from the literature \cite{blasius1907grenzschichten, schlichting2006grenzschicht}.

The adjoint solution employs stored discrete values of the primal procedure. It uses the same discretization of the generalized similarity variable $\eta$ together with a similar shooting approach to compute the adjoint Blasius solution. Using $a=L$ and $b=-1$, the  method aims at hitting the value $\hat{f}^\prime_\infty = 1$ for $\eta \to \infty$ with prescribed wall values $\hat{f}_0 = 0$ and $\hat{f}^{\prime}_0 = 0$ but variable $\hat{f}^{\prime \prime}_0$ for $\eta = 0$. 
Similar to the primal flow, the adjoint boundary value problem is iterated to convergence until $(\hat{f}^\prime - 1) < 10^{-8}$ is reached. 
The numerical results are shown by the right graph of Fig. \ref{fig:blasius_shooting_results} for two simulations that either consider or neglect the ATC term. In case of neglecting ATC, the procedure resembles the primal results. As we will show in the next section, the latter is expected from analytical studies. In case with ATC, the convergence of $\hat{f}^\prime_\infty$ is shifted outwards by one order of magnitude in $\eta$ and results in an increased b.-l. thickness.


%
\section{Analytical Investigation of the Adjoint Blasius Equation}
\label{sec:continuous_blasius_investigation}
Using the similarity transformation introduced in Sec. \ref{sec:adjoint_blasius_equation}, the adjoint b.-l. Eqns. (\ref{equ:adjoint_bounday_layer_mome1_first})-(\ref{equ:adjoint_bounday_layer_mass_first}) were successfully transformed from PDEs into ODEs. Only surface-based objective functional are considered in this paper, for which Eqns. (\ref{equ:tangential_adjoint_blasius_equation})-(\ref{equ:normal_adjoint_blasius_equation}) are generally valid.
The derivation of the continuous adjoint equations reveals the use of Dirichlet conditions for the adjoint velocity along no-slip walls (cf. Tab. \ref{tab:bound_condi}). The latter follows from the objective of interest, which however can occur in different forms depending on the underlying objective functional. 

In the following, we assume either a linear dependence of the functional w.r.t. pressure or no functional along the plate. An illustrative example refers to the evaluation of flow induced forces on the considered plate, where two options to introduce this objective are conceivable. Using an internal formulation, the stresses $[p \delta_{ik} - 2 \nu S_\mathrm{ik}] d_\mathrm{i}$ are first projected in a prescribed spatial direction $d_\mathrm{i}$ and subsequently integrated over (a part of) the plate ($\Delta \Gamma_k$). On the contrary, an external formulation balances the momentum loss between inlet and outlet. From a physical perspective, both approaches pose the same question. However, the adjoint answer is fundamentally different since the habitat of the objective is either the \emph{internal} wall along the plate or the \emph{external} inlet/outlet area. Hence, we end up with different boundary conditions for the wall value, i.e.  $\hat{v}_\mathrm{i} = -d_\mathrm{i}$ or $\hat{v}_\mathrm{i} = 0$ for the interior and the exterior approach respectively. The different approaches resemble the transport theorem formulated in an Arbitrary Lagrangian Eulerian (ALE) frame of reference as indicated in Fig. \ref{fig:adjoint_velocity_profiles} and we refer to K{\"uhl} et al. \cite{kuhl2019decoupling} for a more detailed discussion.
Despite these differences, both formulations employ the similarity transformation to derive a generalized velocity profile, and we can define a normalized adjoint tangential velocity profile by 
\begin{align}
\hat{f}^\prime = \frac{\hat{v}_\mathrm{1}}{\hat{V}_\mathrm{1}}
\qquad \rightarrow \qquad
\hat{f}^\prime = \frac{\hat{v}_\mathrm{1} - \hat{V}_\mathrm{1,w}}{\hat{V}_\mathrm{1,\infty} - \hat{V}_\mathrm{1,w}} \; .
\label{eq:temptt}
\end{align}
Eqn. (\ref{eq:temptt}) reminds of the Blasius solution for a thermal b.-l., e.g. $-2\Theta^{\prime \prime} - \mathrm{Pr} f \Theta^\prime = 0$, where $\Theta = (T - T_\mathrm{w})/(T_\infty - T_\mathrm{w})$ is a non-dimensional temperature and $T$ as well as $\mathrm{Pr}$ represent the temperature and Prandtl number.

\paragraph{Influence of Adjoint Transposed Convection}
Inserting the normal adjoint  Blasius equation (\ref{equ:normal_adjoint_blasius_equation}) into its tangential companion yields the adjoint Blasius Eqn. (\ref{equ:tangential_adjoint_blasius_equation}) without ATC on the right hand side. Choosing a b.-l. measure based on $a=0$ and $b=1$, we can incorporate the primal Blasius expression ($f = -2 f^{\prime \prime \prime} / f^{\prime \prime}$) into the adjoint counterpart and conclude
\begin{align}
\frac{ f^{\prime \prime \prime}}{f^{\prime \prime}} = -\frac{\hat{f}^{\prime \prime \prime}}{\hat{f}^{\prime \prime}}
 \label{equ:conti_inves_witho_advec1}.
\end{align}
We explicitly point out that expression (\ref{equ:conti_inves_witho_advec1}) agrees with an analogue relationship of the thermal b.-l., viz.
\begin{align}
\frac{ f^{\prime \prime \prime}}{f^{\prime \prime}} = \frac{1}{\mathrm{Pr}} \frac{\Theta^{\prime \prime}}{\Theta^\prime} \label{equ:conti_inves_witho_advec} \, , 
\end{align}
for a negative unity Prandtl number $\mathrm{Pr} = -1$, which underlines the reverse flow direction of adjoint systems. Separation of variables allows for a solution for $\hat{f}^\prime$ as well as $\Theta$
\begin{align}
\hat{f}^{\prime} =  C_\mathrm{1} \int_0^{\eta} \frac{1}{f^{\prime \prime}}  \mathrm{d} \eta + C_\mathrm{2}
\qquad \text{and} \qquad
\Theta =  C_\mathrm{3} \int_0^{\eta} \left(f^{\prime \prime}\right)^\mathrm{Pr}  \mathrm{d} \eta + C_\mathrm{4} \;  .
\label{eq:thermalxx}
\end{align}
Unfortunately, the negative unit Prandtl-number of the adjoint system introduces a singularity in (\ref{eq:thermalxx}), since $f^{\prime \prime} \to 0$ for $\eta \to \infty$ holds for the primal system as shown in Sec. \ref{sec:numerical_blasius_approximation}.
The singularity can only be circumvented with an appropriate measure for the adjoint b.-l., viz $a=L$ and $b=-1$. The integration constants of the adjoint velocity profile follow from the boundary conditions for $\hat{f}^\prime (\eta \to 0) = 0$ as well as $\hat{f}^\prime (\eta \to \infty) = 1$ and yield
\begin{align}
C_\mathrm{1} = \int_0^{\infty} f^{\prime \prime}  \mathrm{d} \eta
\qquad \mathrm{and} \qquad
C_\mathrm{2} = 0\; . 
\end{align}
The final solution reads 
\begin{align}
\hat{f}^{\prime}  = \frac{\int_0^{\eta} f^{\prime \prime} \mathrm{d} \eta }{\int_0^{\infty} f^{\prime \prime} \mathrm{d} \eta}
\qquad \longrightarrow \qquad
\hat{f}^{\prime}  = \frac{f^\prime}{f^\prime\left( \infty \right) - f^\prime\left( 0 \right)} = f^\prime.
\end{align}
Hence, the generalized adjoint velocity profile equals its primal counterpart when ATC is suppressed.

\subsection{Estimation of (Dual) Interface Thickness and Wall Shear Stress}
\label{sec:near_plate_investigation}
For known values of the Blasius solution,  
various statements about e.g. b.-l. thickness or shear stress distribution can be derived. This also applies to the adjoint counterparts. Assigning the primal b.-l. thickness to $v_\mathrm{1}/V_\mathrm{1} = 0.99$, yields a value of $\eta_\mathrm{99} \to 4.91 \approx 5$ in line with Fig. \ref{fig:blasius_shooting_results} and Eqn. (\ref{equ:adjoint_similarity})
\begin{align}
\delta_\mathrm{99} \approx 5 \sqrt{\frac{\nu (a+b x_\mathrm{1})}{V_\mathrm{1}}}
\qquad \mathrm{or} \qquad
\frac{\delta_\mathrm{99}}{(a+bx_\mathrm{1})} \approx \frac{5}{\sqrt{\mathrm{Re}_\mathrm{(a+bx_\mathrm{1})}}}.
\end{align}
The numerical and analytical results of the previous sections 
reveal the same b.-l. thickness for primal and adjoint flow, viz. $\hat{\delta}_\mathrm{99} = \delta_\mathrm{99}$ in case of no ATC. However, if the additional right hand side term is incorporated, we end up with approximately $\hat{\eta}_\mathrm{99} \to 5.9424 \approx 6$, cf. Fig. \ref{fig:blasius_shooting_results}, and thus
\begin{align}
\hat{\delta}_\mathrm{99} \approx 6 \sqrt{\frac{\nu (a+b x_\mathrm{1})}{V_\mathrm{1}}}
\qquad \mathrm{or} \qquad
\frac{\hat{\delta}_\mathrm{99}}{(a+bx_\mathrm{1})} \approx \frac{6}{\sqrt{\mathrm{Re}_\mathrm{(a+bx_\mathrm{1})}}}.
\end{align}
%
%
In the remainder of this subsection, the estimated adjoint Blasius values refer to the formulation including ATC, and the index $(\cdot)_\mathrm{w}$ indicates an evaluation along the wall at $\eta = 0$. 
Mind that neglecting the ATC yields strong similarities between the primal and  the dual Blasius solutions, e.g. $\hat{f}_\mathrm{w}^{\prime \prime} = f_\mathrm{w}^{\prime \prime}$ (cf. Sec. \ref{sec:continuous_blasius_investigation}).
Several primal b.-l. thickness measures exist, e.g.   the displacement ($\delta_\mathrm{D}$), momentum ($\delta_\mathrm{M}$) and energy ($\delta_\mathrm{E}$) thickness, which follow from the relation $\mathrm{d} x_\mathrm{2} = \delta \, \mathrm{d} \eta$ 
\begin{alignat}{5}
V_\mathrm{1} \delta_\mathrm{D} &= \int_\mathrm{0}^\mathrm{\infty} \left[ V_\mathrm{1} - v_\mathrm{1}\right] \mathrm{d} x_\mathrm{2}
\qquad &&\longrightarrow \qquad
\delta_\mathrm{D} &&= \int_\mathrm{0}^\mathrm{\infty} \left[1 - f^\prime \right] \mathrm{d} x_\mathrm{2}
\qquad &&\longrightarrow \qquad
\delta_\mathrm{D} &&\approx 1.7208 \sqrt{\frac{\nu (a+b x_\mathrm{1})}{V_\mathrm{1}}} \, , \\
\rho V_\mathrm{1}^2 \delta_\mathrm{M} &= \rho \int_\mathrm{0}^\mathrm{\infty} v_\mathrm{1}\left[ V_\mathrm{1} - v_\mathrm{1}\right] \mathrm{d} x_\mathrm{2}
\qquad &&\longrightarrow \qquad
\delta_\mathrm{M} &&= \int_\mathrm{0}^\mathrm{\infty} f^\prime \left[1 - f^\prime \right] \mathrm{d} x_\mathrm{2}
\qquad &&\longrightarrow \qquad
\delta_\mathrm{M} &&\approx 0.6641 \sqrt{\frac{\nu (a+b x_\mathrm{1})}{V_\mathrm{1}}} \, , \\
\rho V_\mathrm{1}^3 \delta_\mathrm{E} &= \rho \int_\mathrm{0}^\mathrm{\infty} v_\mathrm{1} \left[ V_\mathrm{1}^2 - v_\mathrm{1}^2\right] \mathrm{d} x_\mathrm{2}
\qquad &&\longrightarrow \qquad
\delta_\mathrm{E} &&= \int_\mathrm{0}^\mathrm{\infty} f^\prime\left[1 - {f^\prime}^2 \right] \mathrm{d} x_\mathrm{2}
\qquad &&\longrightarrow \qquad
\delta_\mathrm{E} &&\approx 1.0444 \sqrt{\frac{\nu (a+b x_\mathrm{1})}{V_\mathrm{1}}} \, .
\end{alignat}
 Similar expressions can be derived for the adjoint b.-l., viz.
\begin{alignat}{5}
\hat{V}_\mathrm{1} \hat{\delta}_\mathrm{D} &= \int_\mathrm{0}^\mathrm{\infty} \left[ \hat{V}_\mathrm{1} - \hat{v}_\mathrm{1}\right] \mathrm{d} x_\mathrm{2}
\qquad &&\longrightarrow \qquad
\hat{\delta}_\mathrm{D} &&= \int_\mathrm{0}^\mathrm{1} \left[1 - \hat{f}^\prime \right] \mathrm{d} x_\mathrm{2}
\qquad &&\longrightarrow \qquad
\hat{\delta}_\mathrm{D} &&\approx 2.5336 \sqrt{\frac{\nu (a+b x_\mathrm{1})}{V_\mathrm{1}}} \, , \\
\rho \hat{V}_\mathrm{1}^2 \hat{\delta}_\mathrm{M} &= \rho \int_\mathrm{0}^\mathrm{\infty} \hat{v}_\mathrm{1}\left[ \hat{V}_\mathrm{1} - \hat{v}_\mathrm{1}\right] \mathrm{d} x_\mathrm{2}
\qquad &&\longrightarrow \qquad
\hat{\delta}_\mathrm{M} &&= \int_\mathrm{0}^\mathrm{1} \hat{f}^\prime \left[1 - \hat{f}^\prime \right] \mathrm{d} x_\mathrm{2}
\qquad &&\longrightarrow \qquad
\hat{\delta}_\mathrm{M} &&\approx 0.8430 \sqrt{\frac{\nu (a+b x_\mathrm{1})}{V_\mathrm{1}}}  \, , \\
\rho \hat{V}_\mathrm{1}^3 \hat{\delta}_\mathrm{E} &= \rho \int_\mathrm{0}^\mathrm{\infty} \hat{v}_\mathrm{1} \left[ \hat{V}_\mathrm{1}^2 - \hat{v}_\mathrm{1}^2\right] \mathrm{d} x_\mathrm{2}
\qquad &&\longrightarrow \qquad
\hat{\delta}_\mathrm{E} &&= \int_\mathrm{0}^\mathrm{1} \hat{f}^\prime \left[1 - \hat{f}^{\prime,2} \right] \mathrm{d} x_\mathrm{2}
\qquad &&\longrightarrow \qquad
\hat{\delta}_\mathrm{E} &&\approx 1.2830 \sqrt{\frac{\nu (a+b x_\mathrm{1})}{V_\mathrm{1}}}  \, .
\end{alignat}
The relations between primal and adjoint b.-l. thicknesses read
\begin{align}
\frac{\delta_\mathrm{99}}{\hat{\delta}_\mathrm{99}} \approx 0.8263,
\qquad \qquad
\frac{\delta_\mathrm{D}}{\hat{\delta}_\mathrm{D}} \approx 0.6792,
\qquad \qquad
\frac{\delta_\mathrm{M}}{\hat{\delta}_\mathrm{M}} \approx 0.7878,
\qquad \mathrm{and} \qquad
\frac{\delta_\mathrm{E}}{\hat{\delta}_\mathrm{E}} \approx 0.8140 \, .
\end{align}
Additionally, the generalized Blasius solution offers insight into the primal and adjoint shear stress acting on the plate
\begin{alignat}{4}
\tau_\mathrm{w} &= \mu \frac{\partial v_\mathrm{1}}{\partial x_\mathrm{2}}\bigg|_\mathrm{x_\mathrm{2} = 0} 
&&= \mu V_\mathrm{1} \sqrt{\frac{V_\mathrm{1}}{\nu (a + b x_\mathrm{1})}} f_\mathrm{w}^{\prime \prime} 
\qquad &&\longrightarrow \qquad 
\tau_\mathrm{w} &&\approx 0.3321 \, \mu \, V_\mathrm{1} \sqrt{\frac{V_\mathrm{1}}{\nu (a + b x_\mathrm{1})}} \\
\hat{\tau}_\mathrm{w} &= \mu \frac{\partial \hat{v}_\mathrm{1}}{\partial x_\mathrm{2}}\bigg|_\mathrm{x_\mathrm{2} = 0} 
&&= \mu \hat{V}_\mathrm{1} \sqrt{\frac{V_\mathrm{1}}{\nu (a + b x_\mathrm{1})}} \hat{f}_\mathrm{w}^{\prime \prime} 
\qquad &&\longrightarrow \qquad 
\hat{\tau}_\mathrm{w}  &&\approx 0.1845 \, \mu \, \hat{V}_\mathrm{1} \sqrt{\frac{V_\mathrm{1}}{\nu (a + b x_\mathrm{1})}}.
\end{alignat}
The resulting dual shear is significantly smaller compared to the primal one, e.g. $\tau_\mathrm{w} / \hat{\tau}_\mathrm{w} = f_\mathrm{w}^{\prime \prime} / \hat{f}_\mathrm{w}^{\prime \prime} ( V_\mathrm{1} / \hat{V}_\mathrm{1} )$.
The shear is usually non-dimensionalized via a primal 
($ \rho V_\mathrm{1}^2 / 2$) or an adjoint 
($\rho V_\mathrm{1} \hat{V}_\mathrm{1} / 2$)  dynamic pressure to obtain skin-friction coefficients, i.e.
\begin{alignat}{4}
c_\mathrm{f} &= \frac{\tau_\mathrm{w}}{\frac{1}{2} \rho V_\mathrm{1}^2}
&&= \frac{2 f_\mathrm{w}^{\prime \prime}}{\sqrt{\mathrm{Re}_\mathrm{a + b x_\mathrm{1}}}}
\qquad &&\longrightarrow \qquad
c_\mathrm{f} &&\approx \frac{0.6642}{\sqrt{\mathrm{Re}_\mathrm{a + b x_\mathrm{1}}}} \label{equ:primal_friction_coefficient}  \, \\
\hat{c}_\mathrm{f} &= \frac{\hat{\tau}_\mathrm{w}}{\frac{1}{2} \rho \hat{V}_\mathrm{1} V_\mathrm{1}}
&&= \frac{2 \hat{f}_\mathrm{w}^{\prime \prime}}{\sqrt{\mathrm{Re}_\mathrm{a + b x_\mathrm{1}}}}
\qquad &&\longrightarrow \qquad
\hat{c}_\mathrm{f} &&\approx \frac{0.369}{\sqrt{\mathrm{Re}_\mathrm{a + b x_\mathrm{1}}}}  \label{equ:adjoint_friction_coefficient} \; . 
\end{alignat}
Moreover, known shear-stress distributions allow the integration of a total shear forces on the plate
\begin{alignat}{4}
F_\mathrm{s} &= \int_\mathrm{\Gamma} \tau_\mathrm{w} \mathrm{d} \Gamma_\mathrm{w}
&&= t \int_\mathrm{0}^\mathrm{L} \mu \frac{\partial v_\mathrm{1}}{\partial x_\mathrm{2}}\bigg|_\mathrm{x_\mathrm{2} = 0}  \mathrm{d} x_\mathrm{1}
= 2 f_\mathrm{w}^{\prime \prime} \, t \, V_\mathrm{1} \sqrt{\mu \rho V_\mathrm{1}} \frac{ \sqrt{a + b L} - \sqrt{a}}{b}
\quad &&\longrightarrow \quad
F_\mathrm{s}  &&\approx  0.664 \, t \, V_\mathrm{1} \sqrt{\mu \rho V_\mathrm{1}} \frac{ \sqrt{a + b L} - \sqrt{a}}{b} \label{equ:integrated_primal_drag} \\
\hat{F}_\mathrm{s} &= \int_\mathrm{\Gamma} \hat{\tau}_\mathrm{w} \mathrm{d} \Gamma_\mathrm{w}
&&= t \int_\mathrm{0}^\mathrm{L} \mu \frac{\partial \hat{v}_\mathrm{1}}{\partial x_\mathrm{2}}\bigg|_\mathrm{x_\mathrm{2} = 0}  \mathrm{d} x_\mathrm{1}
= 2 \hat{f}_\mathrm{w}^{\prime \prime} \, t \, \hat{V}_\mathrm{1} \sqrt{\mu \rho V_\mathrm{1}} \frac{ \sqrt{a + b L} - \sqrt{a}}{b}
\quad &&\longrightarrow \quad
\hat{F}_\mathrm{s}  &&\approx  0.369 \, t \, \hat{V}_\mathrm{1} \sqrt{\mu \rho V_\mathrm{1} } \frac{ \sqrt{a + b L} - \sqrt{a}}{b}, \label{equ:integrated_adjoint_drag}
\end{alignat}
where $t$ corresponds to the lateral expansion of the plate. As expected, the choice of the coordinate system has no influence on the forces, since $F_\mathrm{s} (a = 0, b = 1) = F_\mathrm{s} (a = L, b = -1) = 2 f_\mathrm{w}^{\prime \prime} \, t \, \sqrt{\mu \rho L V_\mathrm{1}^3}$ as well as $\hat{F}_\mathrm{s} (a = 0, b = 1) = \hat{F}_\mathrm{s} (a = L, b = -1) = 2 \hat{f}_\mathrm{w}^{\prime \prime} \, t \, \sqrt{\mu \rho L V_\mathrm{1} \hat{V}_\mathrm{1}^2}$. Hence, the ratio between primal and dual shear reads $F_\mathrm{s} / \hat{F}_\mathrm{s} = f_\mathrm{w}^{\prime \prime} / \hat{f}_\mathrm{w}^{\prime \prime} ( V_\mathrm{1} / \hat{V}_\mathrm{1} ) $.
We can compute the primal [adjoint] drag coefficient either from an integration of Eqn. (\ref{equ:primal_friction_coefficient}) [\ref{equ:adjoint_friction_coefficient}] or  from dividing Eqn. (\ref{equ:integrated_primal_drag}) [\ref{equ:integrated_adjoint_drag}] by the dynamic pressure times wetted surface $L \, t$, viz.
\begin{alignat}{5}
c_\mathrm{d} &= \frac{F_\mathrm{s}}{p_\mathrm{\infty} t L} 
&&= \frac{1}{L} \int_\mathrm{0}^\mathrm{L} c_\mathrm{f} \, \mathrm{d} x_\mathrm{1}
&&= \frac{4 f_\mathrm{w}^{\prime \prime}}{\sqrt{Re_\mathrm{L}}} \frac{\sqrt{a + b L} - \sqrt{a}}{b \sqrt{L}}
\qquad &&\longrightarrow \qquad
c_\mathrm{d} &&\approx \frac{1.3284}{\sqrt{Re_\mathrm{L}}}  \frac{\sqrt{a + b L} - \sqrt{a}}{b \sqrt{L}} \label{equ:primal_drag_coefficient} \\
\hat{c}_\mathrm{d} &= \frac{\hat{F}_\mathrm{s}}{\hat{p}_\mathrm{\infty} t L} 
&&= \frac{1}{L} \int_\mathrm{0}^\mathrm{L} \hat{c}_\mathrm{f} \, \mathrm{d} x_\mathrm{1}
&&= \frac{4 \hat{f}_\mathrm{w}^{\prime \prime}}{\sqrt{Re_\mathrm{L}}} \frac{\sqrt{a + b L} - \sqrt{a}}{b \sqrt{L}}
\qquad &&\longrightarrow \qquad
\hat{c}_\mathrm{d} &&\approx \frac{0.738}{\sqrt{Re_\mathrm{L}}}  \frac{\sqrt{a + b L} - \sqrt{a}}{b \sqrt{L}}. \label{equ:adjoint_drag_coefficient}
\end{alignat}
Except different scaling, the dual quantities resemble the well known primal relationship.

According to Eqn. (\ref{equ:shape_derivative}), the combination of primal and dual shear results in a sensitivity distribution along the design surface of the shape. Analogous to the simplification of the shear objective ($J = F_\mathrm{s}$), the approach also applies to its shape sensitivity, viz.
\begin{align}
s = - \nu \frac{\tau_\mathrm{w}}{\mu} \frac{\hat{\tau}_\mathrm{w}}{\mu}
= - \nu \frac{\partial v_\mathrm{1}}{\partial x_\mathrm{2}}\bigg|_\mathrm{x_\mathrm{2} = 0}  \frac{\partial \hat{v}_\mathrm{1}}{\partial x_\mathrm{2}}\bigg|_\mathrm{x_\mathrm{2} = 0}
= -f_\mathrm{w}^{\prime \prime} \hat{f}_\mathrm{w}^{\prime \prime} \frac{\hat{V}_\mathrm{1} V_\mathrm{1}^2}{a + b x_\mathrm{1}}
\qquad \longrightarrow \qquad
s \approx -0.06127 \frac{ \hat{V}_\mathrm{1} V_\mathrm{1}^2}{a + b x_\mathrm{1}} \label{equ:simplified_sensitivity_local}.
\end{align}
The local shape derivative has again singularities at the leading and trailing edge of the plate. However, the intermediate behavior scales with $x^{-1}$ instead of $x^{-1/2}$.
Furthermore, the local sensitivity increases quadratically with the primal but only linearly with the adjoint velocity. This reminds of the quadratic character of the primal NS equations (e.g. $v_\mathrm{k} \partial v_\mathrm{i} / \partial x_\mathrm{k}$) which are opposed by the linear nature of the adjoint counterpart (e.g. $v_\mathrm{k} \partial \hat{v}_\mathrm{i} / \partial x_\mathrm{k}$).
The sensitivity expression (\ref{equ:simplified_sensitivity_local}) can be non-dimensionalized towards a sensitivity-coefficient $c_\mathrm{s}$ via the kinematic viscosity and the primal as well as dual dynamic pressure or by combining the primal and adjoint skin-friction coefficient, viz.
\begin{align}
c_\mathrm{s} 
= s \frac{\rho^2 \nu}{p_\mathrm{\infty} \hat{p}_\mathrm{\infty}} 
= s \frac{4  \nu}{ V_\mathrm{1}^3 \hat{V}_\mathrm{1}}
= -  c_\mathrm{f} \, \hat{c}_\mathrm{f} 
= - \frac{ 4 f_\mathrm{w}^{\prime \prime} \hat{f}_\mathrm{w}^{\prime \prime}}{\mathrm{Re}_\mathrm{a + b x_\mathrm{1}}}
\qquad \longrightarrow \qquad
c_\mathrm{s} \approx -\frac{0.06127}{\mathrm{Re}_\mathrm{a + b x_\mathrm{1}} } \label{equ:sensitivity_coefficient}.
\end{align}
Finally, the integration of (\ref{equ:simplified_sensitivity_local}) along the plate provides the sensitivity derivative on an integral level
\begin{align}
\delta_\mathrm{u} J 
= t \, \int_0^L s \, \mathrm{d} x_\mathrm{1}
= -f_\mathrm{w}^{\prime \prime} \hat{f}_\mathrm{w}^{\prime \prime} \frac{t \hat{V}_\mathrm{1} V_\mathrm{1}^2}{b} \mathrm{ln}\left[\frac{a+bL}{a}\right]
\qquad \longrightarrow \qquad
\delta_\mathrm{u} J \approx - 0.06127 \frac{t \hat{V}_\mathrm{1} V_\mathrm{1}^2}{b} \mathrm{ln}\left[\frac{a+bL}{a}\right] \label{equ:simplified_sensitivity_global}.
\end{align}
Interestingly, a singularity arises for the integral sensitivity. Owing to the proportionality $s \propto x^{-1}$, 
a logarithmic $\delta_\mathrm{u} J \propto \pm (\mathrm{ln}(L) - \mathrm{ln}(0))$ relationship results for the integral sensitivity.
The $s \propto x^{-1}$ relationship originates in the definition of the similarity variable which in turn estimates the primal interface thickness by $\delta \propto x^{1/2}$ that finally yields $s \propto 1/\delta^2$.
The singularity cannot be avoided by adjusting the coordinate system (i.e. $a$ and $b$), since the integral bounds also need to be adjusted to $a$ and $b$. Mind that we apply only $a=0,b=1$ or $a=L,b=-1$ in this manuscript.
It seems that the plate has an infinite potential to reduce its flow resistance from an integral point of view. This statement seems suspicious at first. However, mind that a disappearing plate would wipe out its resistance completely. 
A perturbation into the plate normal affects the drag via a variation in the local shear $\delta F \propto \delta f_\mathrm{w}^{\prime \prime}$ (cf. Eqn. (\ref{equ:integrated_primal_drag})) which in turn follows from a variation in the similarity variable $\delta f_\mathrm{w}^{\prime \prime} \to f_\mathrm{w}^{\prime \prime \prime} \delta \eta $ that finally can be estimated via $\delta \eta \propto \delta x_\mathrm{2} / \sqrt{x_\mathrm{1}} - 0.5 (\eta / x_\mathrm{1}) \delta x_\mathrm{1}$ for an interface measure that employs $a = 0$ and $b = 1$. Thus, $\delta F \propto f_\mathrm{w}^{\prime \prime \prime}[\delta x_\mathrm{2} / \sqrt{x_\mathrm{1}} - 0.5 (\eta / x_\mathrm{1}) \delta x_\mathrm{1}]$ becomes singular if both coordinates tend to zero. 
Note that the drag as well as the similarity variable become singular at the leading edge and the influence of a plate-normal variation is debatable, not least because the direction of the normal is not defined at the leading edge.

The integral shape derivative can be non-dimensionalized based on the dynamic pressures and the wetted surface.
An alternative approach to a global sensitivity coefficient follows from the  
integration over the sensitivity coefficient, viz. 
\begin{align}
c_\mathrm{\delta_\mathrm{u} J}
= \frac{\delta_\mathrm{u} J \, \nu}{ p_\mathrm{\infty} \hat{p}_\mathrm{\infty} t L}
= - \frac{1}{L} \int_\mathrm{0}^\mathrm{L} c_\mathrm{s} \, \mathrm{d} x_\mathrm{1}
= - \frac{ f_\mathrm{w}^{\prime \prime} \hat{f}_\mathrm{w}^{\prime \prime}}{\mathrm{Re}_\mathrm{L}}  \frac{1}{b} \mathrm{ln}\left[\frac{a+bL}{a}\right]
\qquad \longrightarrow \qquad
c_\mathrm{\delta_\mathrm{u} J} \approx - \frac{0.06127}{\mathrm{Re}_\mathrm{L}}  \frac{1}{b} \mathrm{ln}\left[\frac{a+bL}{a}\right] \label{equ:simplified_sensitivity_coefficient_global}.
\end{align}
However, we would like to point out that an interpretation of adjoint values, regarding e.g. units or the definition of a sensitivity rule, depends strongly on the objective under investigation. The same holds for the definition of the control.

%
\section{Applications}
\label{sec:application}
\begin{figure}
\centering
\input{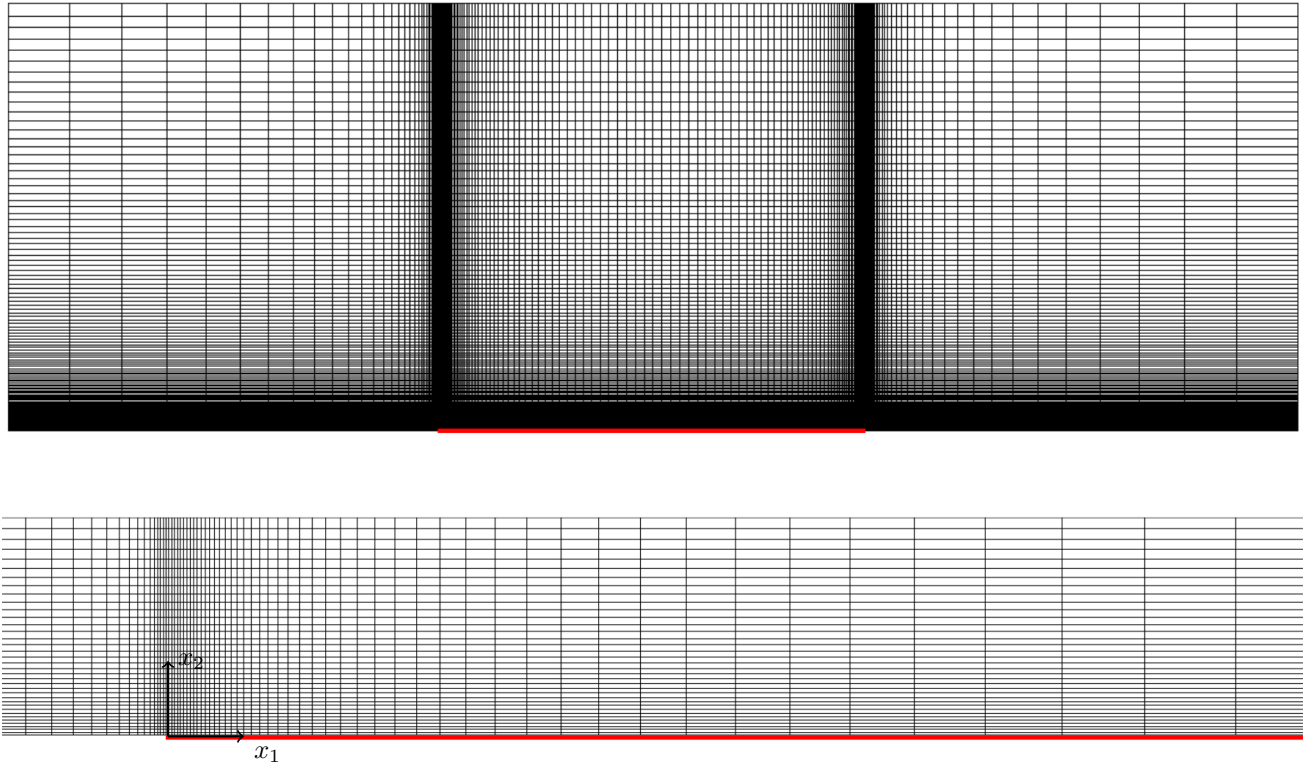}
\caption{Employed structured grid for the flat plate flow. 
Complete domain (top) and  refined region around the leading edge where the origin of the coordinate system is located (bottom). The plate is indicated by red lines. }
\label{fig:plate_grid}
\end{figure}

This section aims at the numerical assessment of the theoretical observations reported in the previous chapters. For this reason the flow over a simple, finite-length flat plate is considered, cf. Figs. \ref{fig:plate_flow} and \ref{fig:plate_grid}.
The computational studies are performed for laminar flows at Reynolds-numbers between $10^3 \leq \mathrm{Re}_\mathrm{L} = V_\mathrm{1} L/\nu \leq 10^5$ based on a homogeneous flow velocity $V_\mathrm{1}$ and the kinematic viscosity $\nu$ using a Navier-Stokes procedure.

\paragraph{Numerical Procedure}
The numerical procedure is based upon the Finite-Volume procedure FreSCo+ \cite{rung2009challenges}. Analogue to the use of integration-by-parts in deriving the continuous adjoint equations, summation-by-parts is employed to derive the building blocks of the discrete (dual) adjoint expressions. A detailed derivation of this hybrid adjoint approach can be found in \cite{stuck2013adjoint, kroger2018adjoint, kuhl2020adjoint}. The segregated algorithm uses a cell-centered, collocated storage arrangement for all transport properties. The implicit numerical approximation is second order accurate and supports polyhedral cells. Both, the primal and adjoint pressure-velocity coupling is based on a SIMPLE method and possible parallelization is realized by means of a domain decomposition approach \cite{yakubov2013hybrid, yakubov2015experience}. In terms of a CAD-free shape optimisation approach, the computational grid can be adjusted using a Laplace-Beltrami \cite{stuck2013adjoint,kroger2015cad} or Steklov-Poincar\'e \cite{schulz2016computational, kuhl2020adjoint} type (surface metric) approach.

\paragraph{Computational Model and Numerical Grid} The two-dimensional domain has a length and height of $3 \, L$ and $L$. The inlet and top boundaries are located one length away from the origin of the coordinate system, cf. Fig. \ref{fig:plate_grid}. The latter is located in the leading edge of the plate.
The velocity is prescribed at the inlet, a slip wall is used along the top boundary and symmetry conditions are employed along the bottom before and after the plate. Zero gradient conditions are employed at the outlet.
The convective term for primal [adjoint] momentum is approximated using the Quadratic Upstream [Downstream] Interpolation of Convective Kinematics (QUICK) [QU(D)ICK] scheme.

To ensure the independence of the objective functional w.r.t. spatial discretization, a grid study was conducted.
The considered five grids are all symmetric w.r.t. the mid-plate at $x_1=L/2$, and the grid spacing in horizontal and vertical direction was successively halved between two consecutive grids.
Figure \ref{fig:plate_primal_integral_results} depicts the evolution of the drag coefficient $c_\mathrm{d}$ over the grid refinement level -- indicated by the number of control volumes $n_\mathrm{fv}$ -- for 
 an exemplary flow at $\mathrm{Re}_\mathrm{L} = 10^4$.
Based on the grid convergence studies, all results shown hereafter were obtained for the finest grid level that consists of approximately $28.000$ control volumes. The controlled plate shape is discretized with 160 surface elements and the boundary layers were typically resolved by more than 50 control volumes. Since the investigated Reynolds numbers vary by two orders of magnitude, the plate normal resolution was adjusted to ensure $y^\mathrm{+} = \mathcal{O}(10^{-1})$ in all cases. 

\subsection{Primal Flow Results}
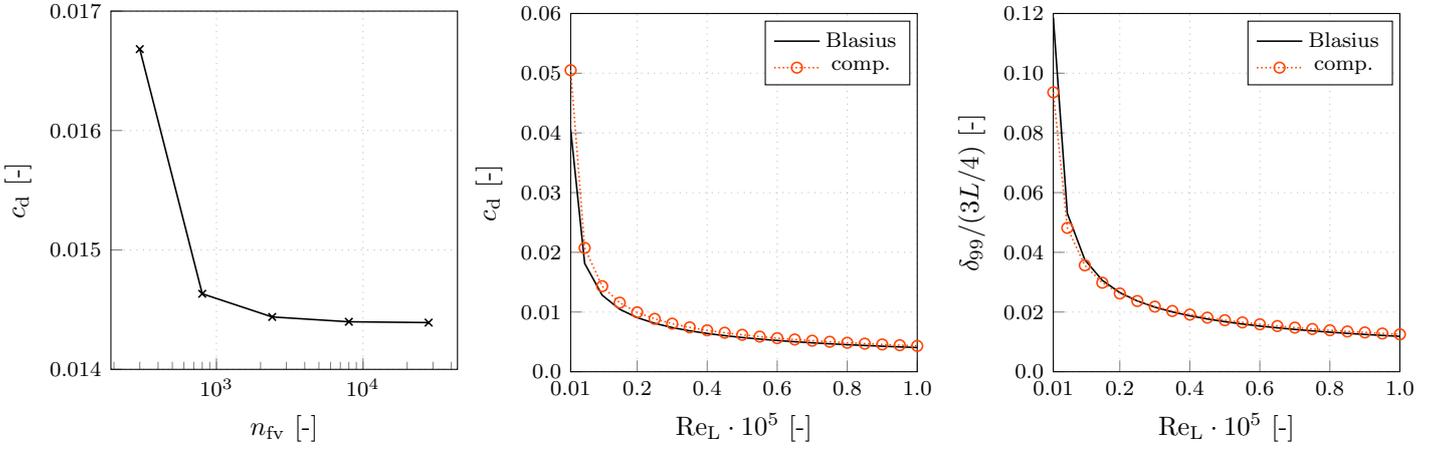
\begin{figure}
\centering
\analytiSolutionPictures
\begin{tikzpicture}
\begin{axis}[
 ylabel style={text width=0.25\textwidth,align=center},
 xlabel={$n_\mathrm{fv}$ [-]},
 ylabel={$c_\mathrm{d}$ [-]},
 ymin=0.014,ymax=0.017,
 xmode = log,
 ytick={0.014, 0.015,0.016,0.017},
 yticklabels={0.014, 0.015,0.016,0.017},
 scaled y ticks = false
]

\addplot [line1, mark1] table[x expr={\thisrowno{0}},y expr={\thisrowno{1}}] {data/Plate_Grid_Study.dat};
 
\end{axis}
\end{tikzpicture}
\begin{tikzpicture}
\begin{axis}[
 ylabel style={text width=0.25\textwidth,align=center},
 xlabel={$\mathrm{Re}_\mathrm{L} \cdot 10^5$ [-]},
 ylabel={$c_\mathrm{d}$ [-]},
 xmin=1000,xmax=100000,
 ymin=0.0,ymax=0.06,
 xtick={1000,20000,40000,60000,80000,100000},
 xticklabels={0.01,0.2,0.4,0.6,0.8,1.0},
 scaled x ticks = false,
 ytick={0.0,0.01,0.02,0.03,0.04,0.05,0.06},
 yticklabels={0.0,0.01,0.02,0.03,0.04,0.05,0.06},
 scaled y ticks = false
]
\addplot [line1] table[x expr={\thisrowno{0}},y expr={\thisrowno{1}}] {data/Plate_Primal_Drag_Coefficient.dat};
\addplot [line2, mark2] table[x expr={\thisrowno{0}},y expr={\thisrowno{2}}] {data/Plate_Primal_Drag_Coefficient.dat};
 
\addlegendentry{Blasius};
\addlegendentry{comp.};

\end{axis}
\end{tikzpicture}
\begin{tikzpicture}
\begin{axis}[
 ylabel style={text width=0.25\textwidth,align=center},
 xlabel={$\mathrm{Re}_\mathrm{L} \cdot 10^5$ [-]},
 ylabel={$\delta_\mathrm{99} / (3L/4)$ [-]},
 xmin=1000,xmax=100000,
 ymin=0.0,ymax=0.12,
 xtick={1000,20000,40000,60000,80000,100000},
 xticklabels={0.01,0.2,0.4,0.6,0.8,1.0},
 scaled x ticks = false,
 ytick={0.0,0.02,0.04,0.06,0.08,0.10,0.12},
 yticklabels={0.0,0.02,0.04,0.06,0.08,0.10,0.12},
 scaled y ticks = false
]
\addplot [line1] table[x expr={\thisrowno{0}},y expr={\thisrowno{1}}] {data/Plate_Primal_Integral_Thicknesses.dat};
\addplot [line2, mark2] table[x expr={\thisrowno{0}},y expr={\thisrowno{2}}] {data/Plate_Primal_Integral_Thicknesses.dat};
 
\addlegendentry{Blasius};
\addlegendentry{comp.};

\end{axis}
\end{tikzpicture}
\caption{Predicted integral parameters for the primal flat plate flow. Evolution of the frag coefficient $c_\mathrm{d}$ over the grid refinement level -- indicated by the amount of control volumes -- $n_\mathrm{fv}$ (left), comparison of Blasius and numerical  drag coefficient as well as b.-l. thickness ($\delta_\mathrm{99}$) predictions  at $x/L = 3/4$ for a range of Reynolds numbers  (center, right).}
\label{fig:plate_primal_integral_results}
\end{figure}

The quality of adjoint results hinges on the quality of the preceding primal approximation. Therefore, first the predictive accuracy of the estimates made in Sec. \ref{sec:near_plate_investigation} is compared against primal Navier-Stokes results. 
%
Fig. \ref{fig:plate_primal_integral_results} compares the predicted results for drag coefficient $c_\mathrm{d}$ (center) and 99\%-b.-l. thickness $\delta_\mathrm{99}$ at $x_\mathrm{1} / L = 3 / 4$ (right) against the Blasius results based on a measure that employs $a=0$ and $b=1$. 
Quantitatively, the resistance coefficient  [interface thickness] obtained from a Blasius solution is slightly overestimated [underestimated] for small Reynolds numbers, whereby the qualitative behavior is still in fair agreement.

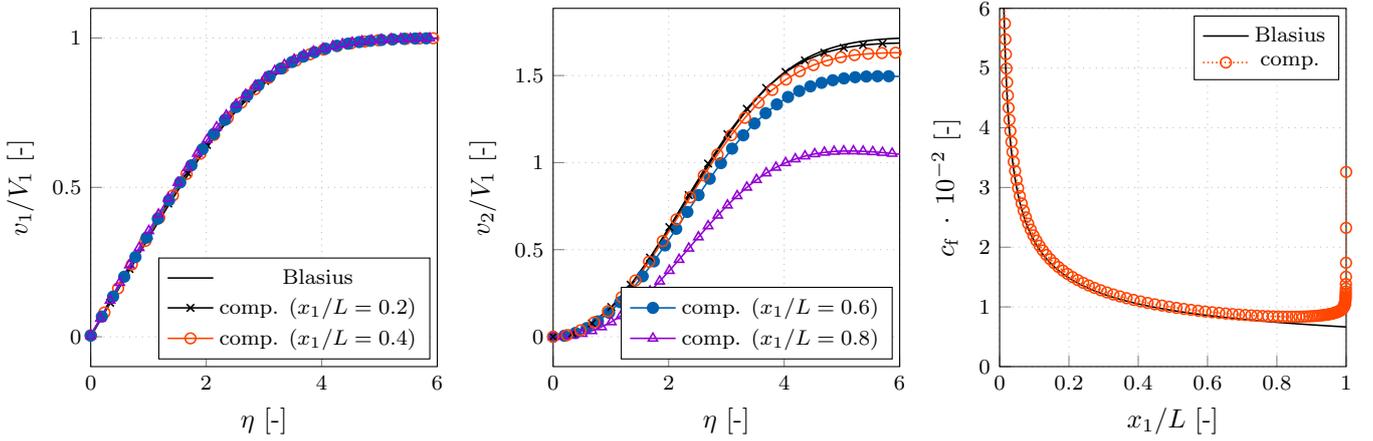
\begin{figure}
\centering
\analytiSolutionPictures
\begin{tikzpicture}
\begin{axis}[
 ylabel style={text width=0.25\textwidth,align=center},
 xlabel={$\eta$ [-]},
 ylabel={$v_\mathrm{1} / V_\mathrm{1}$ [-]},
 legend style={at={(0.98,0.02)},anchor=south east},
 xmin=0,xmax=6,
]

\addplot [line1] table[x expr={\thisrowno{0}},y expr={\thisrowno{2}}] {data/Blasius_Shooting_Primal_Results.dat};
\addplot [mark1] table[x expr={\thisrowno{2}},y expr={\thisrowno{3}}] {data/Plate_Primal_Blasius_FreSCo_V1_Results.dat};
\addplot [mark2] table[x expr={\thisrowno{6}},y expr={\thisrowno{7}}] {data/Plate_Primal_Blasius_FreSCo_V1_Results.dat};
\addplot [mark3] table[x expr={\thisrowno{10}},y expr={\thisrowno{11}}] {data/Plate_Primal_Blasius_FreSCo_V1_Results.dat};
\addplot [mark4] table[x expr={\thisrowno{14}},y expr={\thisrowno{15}}] {data/Plate_Primal_Blasius_FreSCo_V1_Results.dat};


\addlegendentry{Blasius};
\addlegendentry{comp. ($x_\mathrm{1}/L = 0.2$)};
\addlegendentry{comp. ($x_\mathrm{1}/L = 0.4$)};
 
\end{axis}
\end{tikzpicture}
\begin{tikzpicture}
\begin{axis}[
 ylabel style={text width=0.25\textwidth,align=center},
 xlabel={$\eta$ [-]},
 ylabel={$v_\mathrm{2} / V_\mathrm{1}$ [-]},
 legend style={at={(0.98,0.02)},anchor=south east},
 xmin=0,xmax=6,
]

\addplot [mark3] table[x expr={\thisrowno{10}},y expr={\thisrowno{11}}] {data/Plate_Primal_Blasius_FreSCo_V2_Results.dat};
\addplot [mark4] table[x expr={\thisrowno{14}},y expr={\thisrowno{15}}] {data/Plate_Primal_Blasius_FreSCo_V2_Results.dat};
\addplot [line1] table[x expr={\thisrowno{0}},y expr={\thisrowno{4}}] {data/Blasius_Shooting_Primal_Results.dat};
\addplot [mark1] table[x expr={\thisrowno{2}},y expr={\thisrowno{3}}] {data/Plate_Primal_Blasius_FreSCo_V2_Results.dat};
\addplot [mark2] table[x expr={\thisrowno{6}},y expr={\thisrowno{7}}] {data/Plate_Primal_Blasius_FreSCo_V2_Results.dat};


\addlegendentry{comp. ($x_\mathrm{1}/L = 0.6$)};
\addlegendentry{comp. ($x_\mathrm{1}/L = 0.8$)};
 
\end{axis}
\end{tikzpicture}
\begin{tikzpicture}
\begin{axis}[
 ylabel style={text width=0.25\textwidth,align=center},
 xlabel={$x_\mathrm{1} / L$ [-]},
 ylabel={$c_\mathrm{f} \cdot 10^{-2}$ [-]},
 xmin=0.0,xmax=1.0,
 ymin=0.0,ymax=0.06,
 ytick={0.0,0.01,0.02,0.03,0.04,0.05,0.06},
 yticklabels={0,1,2,3,4,5,6},
 scaled y ticks = false
]
\addplot [line1] table[x expr={\thisrowno{0}},y expr={\thisrowno{2}}] {data/Blasius_Shooting_Primal_Friction_Coefficient.dat};
\addplot [line2, mark2] table[x expr={\thisrowno{0}},y expr={\thisrowno{1}}] {data/Blasius_Shooting_Primal_Friction_Coefficient.dat};
 
\addlegendentry{Blasius};
\addlegendentry{comp.};

\end{axis}
\end{tikzpicture}
\caption{Comparison of Blasius and Navier-Stokes results for the primal flow over a flat plate at $\mathrm{Re}_\mathrm{L} = 10^4$. Tangential velocity ($v_\mathrm{1} / V_\mathrm{1}$) against $f^\prime$ (left) and  normal velocity ($v_\mathrm{2} / V_\mathrm{1}$) against $f^\prime \eta - f$ (center) at four different locations. Skin-friction coefficient $c_\mathrm{f}$ against $0.664 / \mathrm{Re}_{\mathrm{x}_\mathrm{1}}$ based on $a=0$ and $b=1$ (right).}
\label{fig:plate_primal_local_results}
\end{figure}

Supplementary to the
comparison of integral parameters depicted by Fig. \ref{fig:plate_primal_integral_results}, local results were examined. 
Results displayed in the present paper  are confined to $\mathrm{Re}_\mathrm{L} = 10^4$. Figure  \ref{fig:plate_primal_local_results} compares the normalized tangential $v_\mathrm{1}/V_\mathrm{1}$ (left) and normal $v_\mathrm{2}/V_\mathrm{1}$ (center) velocity profile against the similarity solution for a range of locations. 
While the tangential velocity  fits almost perfectly with the Blasius solution, the normal component deviates significantly from the expected solution when the b.-l. approaches the trailing edge. This phenomenon is attributed to the abrupt change of the boundary condition for the examined finite-length plate, cf. Fig. \ref{fig:plate_grid}, and the ability of trailing-edge information to propagate upstream in a Navier-Stokes framework. 
%
As also shown in Fig. \ref{fig:plate_primal_local_results}, pronounced deviations from Blasius solutions occur for the local skin-friction coefficient $c_\mathrm{f}$ at the trailing edge and approximately 20\% upstream. 

\subsection{Adjoint Flow Results}
The adjoint investigations were performed for a drag objective ($d_\mathrm{i} = [1, 0]^\mathrm{T}$) on the basis of two different formulations referred to as interior (FI) or exterior (FE) drag-force evaluation, cf. section \ref{sec:continuous_blasius_investigation}.
Additionally, we distinguish between simulations that neglect (A0) or employ (A1) ATC. Thus,  four adjoint computations were conducted for each primal flow.

\begin{figure}
\centering
\longPicture
\pgfplotsset{
	legend columns=3
}
\begin{tikzpicture}
\begin{axis}[
 xlabel style={text width=0.25\textwidth,align=center},
 ylabel style={text width=0.25\textwidth,align=center},
 xlabel={$x_\mathrm{1} / L$ [-]},
 ylabel={$x_\mathrm{2} / L$ [-]},
 xmin=-0.01,xmax=1.01,
 ymin=-0.0,ymax=0.061,
 ytick={0.0, 0.01, 0.02, 0.03, 0.04, 0.05, 0.06},
 yticklabels={0.0, 0.01, 0.02, 0.03, 0.04, 0.05, 0.06},
 scaled y ticks = false
]

\addplot [line3] table[x expr={\thisrowno{0}}, y expr={\thisrowno{2}}] {data/Plate_Adjoint_Profile_Thickness_Blasius.dat};
\addplot [line5] table[x expr={\thisrowno{0}}, y expr={\thisrowno{1}}] {data/Plate_Adjoint_Profile_Thickness_External.dat};
\addplot [line1] table[x expr={\thisrowno{2}}, y expr={\thisrowno{1}}] {data/Plate_Adjoint_Profile_External_Pos_0.dat};

\addplot [line4] table[x expr={\thisrowno{0}}, y expr={\thisrowno{1}}] {data/Plate_Adjoint_Profile_Thickness_Blasius.dat};
\addplot [line6] table[x expr={\thisrowno{0}}, y expr={\thisrowno{1}}] {data/Plate_Adjoint_Profile_Thickness_External_Without_Advection.dat};
\addplot [line2] table[x expr={\thisrowno{4}}, y expr={\thisrowno{5}}] {data/Plate_Adjoint_Profile_Internal_Pos_0.dat};


\addplot [line1] table[x expr={\thisrowno{0}}, y expr={\thisrowno{1}}] {data/Plate_Adjoint_Profile_External_Pos_0.dat};
\addplot [line1, quiver={u=\thisrowno{2},v=\thisrowno{3}},-stealth,each nth point=1] table {data/Plate_Adjoint_Profile_External_Pos_0.dat};

\addplot [line1] table[x expr={\thisrowno{0}}, y expr={\thisrowno{1}}] {data/Plate_Adjoint_Profile_External_Pos_1.dat};
\addplot [line1] table[x expr={\thisrowno{4}}, y expr={\thisrowno{5}}] {data/Plate_Adjoint_Profile_External_Pos_1.dat};
\addplot [line1, quiver={u=\thisrowno{2},v=\thisrowno{3}},-stealth,each nth point=1] table {data/Plate_Adjoint_Profile_External_Pos_1.dat};

\addplot [line1] table[x expr={\thisrowno{0}}, y expr={\thisrowno{1}}] {data/Plate_Adjoint_Profile_External_Pos_2.dat};
\addplot [line1] table[x expr={\thisrowno{4}}, y expr={\thisrowno{5}}] {data/Plate_Adjoint_Profile_External_Pos_2.dat};
\addplot [line1, quiver={u=\thisrowno{2},v=\thisrowno{3}},-stealth,each nth point=1] table {data/Plate_Adjoint_Profile_External_Pos_2.dat};

\addplot [line1] table[x expr={\thisrowno{0}}, y expr={\thisrowno{1}}] {data/Plate_Adjoint_Profile_External_Pos_3.dat};
\addplot [line1] table[x expr={\thisrowno{4}}, y expr={\thisrowno{5}}] {data/Plate_Adjoint_Profile_External_Pos_3.dat};
\addplot [line1, quiver={u=\thisrowno{2},v=\thisrowno{3}},-stealth,each nth point=1] table {data/Plate_Adjoint_Profile_External_Pos_3.dat};

\addplot [line1] table[x expr={\thisrowno{0}}, y expr={\thisrowno{1}}] {data/Plate_Adjoint_Profile_External_Pos_4.dat};
\addplot [line1] table[x expr={\thisrowno{4}}, y expr={\thisrowno{5}}] {data/Plate_Adjoint_Profile_External_Pos_4.dat};
\addplot [line1, quiver={u=\thisrowno{2},v=\thisrowno{3}},-stealth,each nth point=1] table {data/Plate_Adjoint_Profile_External_Pos_4.dat};

\addplot [line1] table[x expr={\thisrowno{0}}, y expr={\thisrowno{1}}] {data/Plate_Adjoint_Profile_External_Pos_5.dat};
\addplot [line1] table[x expr={\thisrowno{4}}, y expr={\thisrowno{5}}] {data/Plate_Adjoint_Profile_External_Pos_5.dat};
\addplot [line1, quiver={u=\thisrowno{2},v=\thisrowno{3}},-stealth,each nth point=1] table {data/Plate_Adjoint_Profile_External_Pos_5.dat};

\addplot [line1] table[x expr={\thisrowno{0}}, y expr={\thisrowno{1}}] {data/Plate_Adjoint_Profile_External_Pos_6.dat};
\addplot [line1] table[x expr={\thisrowno{4}}, y expr={\thisrowno{5}}] {data/Plate_Adjoint_Profile_External_Pos_6.dat};
\addplot [line1, quiver={u=\thisrowno{2},v=\thisrowno{3}},-stealth,each nth point=1] table {data/Plate_Adjoint_Profile_External_Pos_6.dat};

\addplot [line1] table[x expr={\thisrowno{0}}, y expr={\thisrowno{1}}] {data/Plate_Adjoint_Profile_External_Pos_7.dat};
\addplot [line1] table[x expr={\thisrowno{4}}, y expr={\thisrowno{5}}] {data/Plate_Adjoint_Profile_External_Pos_7.dat};
\addplot [line1, quiver={u=\thisrowno{2},v=\thisrowno{3}},-stealth,each nth point=1] table {data/Plate_Adjoint_Profile_External_Pos_7.dat};

\addplot [line1] table[x expr={\thisrowno{0}}, y expr={\thisrowno{1}}] {data/Plate_Adjoint_Profile_External_Pos_8.dat};
\addplot [line1] table[x expr={\thisrowno{4}}, y expr={\thisrowno{5}}] {data/Plate_Adjoint_Profile_External_Pos_8.dat};
\addplot [line1, quiver={u=\thisrowno{2},v=\thisrowno{3}},-stealth,each nth point=1] table {data/Plate_Adjoint_Profile_External_Pos_8.dat};

\addplot [line1] table[x expr={\thisrowno{0}}, y expr={\thisrowno{1}}] {data/Plate_Adjoint_Profile_External_Pos_9.dat};
\addplot [line1] table[x expr={\thisrowno{4}}, y expr={\thisrowno{5}}] {data/Plate_Adjoint_Profile_External_Pos_9.dat};
\addplot [line1, quiver={u=\thisrowno{2},v=\thisrowno{3}},-stealth,each nth point=1] table {data/Plate_Adjoint_Profile_External_Pos_9.dat};



\addplot [line2] table[x expr={\thisrowno{0}}, y expr={\thisrowno{1}}] {data/Plate_Adjoint_Profile_Internal_Pos_0.dat};
\addplot [line2, quiver={u=\thisrowno{2},v=\thisrowno{3}},-stealth,each nth point=1] table {data/Plate_Adjoint_Profile_Internal_Pos_0.dat};

\addplot [line2] table[x expr={\thisrowno{0}}, y expr={\thisrowno{1}}] {data/Plate_Adjoint_Profile_Internal_Pos_1.dat};
\addplot [line2] table[x expr={\thisrowno{4}}, y expr={\thisrowno{5}}] {data/Plate_Adjoint_Profile_Internal_Pos_1.dat};
\addplot [line2, quiver={u=\thisrowno{2},v=\thisrowno{3}},-stealth,each nth point=1] table {data/Plate_Adjoint_Profile_Internal_Pos_1.dat};

\addplot [line2] table[x expr={\thisrowno{0}}, y expr={\thisrowno{1}}] {data/Plate_Adjoint_Profile_Internal_Pos_2.dat};
\addplot [line2] table[x expr={\thisrowno{4}}, y expr={\thisrowno{5}}] {data/Plate_Adjoint_Profile_Internal_Pos_2.dat};
\addplot [line2, quiver={u=\thisrowno{2},v=\thisrowno{3}},-stealth,each nth point=1] table {data/Plate_Adjoint_Profile_Internal_Pos_2.dat};

\addplot [line2] table[x expr={\thisrowno{0}}, y expr={\thisrowno{1}}] {data/Plate_Adjoint_Profile_Internal_Pos_3.dat};
\addplot [line2] table[x expr={\thisrowno{4}}, y expr={\thisrowno{5}}] {data/Plate_Adjoint_Profile_Internal_Pos_3.dat};
\addplot [line2, quiver={u=\thisrowno{2},v=\thisrowno{3}},-stealth,each nth point=1] table {data/Plate_Adjoint_Profile_Internal_Pos_3.dat};

\addplot [line2] table[x expr={\thisrowno{0}}, y expr={\thisrowno{1}}] {data/Plate_Adjoint_Profile_Internal_Pos_4.dat};
\addplot [line2] table[x expr={\thisrowno{4}}, y expr={\thisrowno{5}}] {data/Plate_Adjoint_Profile_Internal_Pos_4.dat};
\addplot [line2, quiver={u=\thisrowno{2},v=\thisrowno{3}},-stealth,each nth point=1] table {data/Plate_Adjoint_Profile_Internal_Pos_4.dat};

\addplot [line2] table[x expr={\thisrowno{0}}, y expr={\thisrowno{1}}] {data/Plate_Adjoint_Profile_Internal_Pos_5.dat};
\addplot [line2] table[x expr={\thisrowno{4}}, y expr={\thisrowno{5}}] {data/Plate_Adjoint_Profile_Internal_Pos_5.dat};
\addplot [line2, quiver={u=\thisrowno{2},v=\thisrowno{3}},-stealth,each nth point=1] table {data/Plate_Adjoint_Profile_Internal_Pos_5.dat};

\addplot [line2] table[x expr={\thisrowno{0}}, y expr={\thisrowno{1}}] {data/Plate_Adjoint_Profile_Internal_Pos_6.dat};
\addplot [line2] table[x expr={\thisrowno{4}}, y expr={\thisrowno{5}}] {data/Plate_Adjoint_Profile_Internal_Pos_6.dat};
\addplot [line2, quiver={u=\thisrowno{2},v=\thisrowno{3}},-stealth,each nth point=1] table {data/Plate_Adjoint_Profile_Internal_Pos_6.dat};

\addplot [line2] table[x expr={\thisrowno{0}}, y expr={\thisrowno{1}}] {data/Plate_Adjoint_Profile_Internal_Pos_7.dat};
\addplot [line2] table[x expr={\thisrowno{4}}, y expr={\thisrowno{5}}] {data/Plate_Adjoint_Profile_Internal_Pos_7.dat};
\addplot [line2, quiver={u=\thisrowno{2},v=\thisrowno{3}},-stealth,each nth point=1] table {data/Plate_Adjoint_Profile_Internal_Pos_7.dat};

\addplot [line2] table[x expr={\thisrowno{0}}, y expr={\thisrowno{1}}] {data/Plate_Adjoint_Profile_Internal_Pos_8.dat};
\addplot [line2] table[x expr={\thisrowno{4}}, y expr={\thisrowno{5}}] {data/Plate_Adjoint_Profile_Internal_Pos_8.dat};
\addplot [line2, quiver={u=\thisrowno{2},v=\thisrowno{3}},-stealth,each nth point=1] table {data/Plate_Adjoint_Profile_Internal_Pos_8.dat};

\addplot [line2] table[x expr={\thisrowno{0}}, y expr={\thisrowno{1}}] {data/Plate_Adjoint_Profile_Internal_Pos_9.dat};
\addplot [line2] table[x expr={\thisrowno{4}}, y expr={\thisrowno{5}}] {data/Plate_Adjoint_Profile_Internal_Pos_9.dat};
\addplot [line2, quiver={u=\thisrowno{2},v=\thisrowno{3}},-stealth,each nth point=1] table {data/Plate_Adjoint_Profile_Internal_Pos_9.dat};


\addlegendentry{Blas. $\hat{\delta}$ (A1)};
\addlegendentry{comp. $\hat{\delta}$ (A1)}; 
\addlegendentry{FE, A1}; 

\addlegendentry{Blas. $\hat{\delta}$ (A0)}; 
\addlegendentry{comp. $\hat{\delta}$ (A0)}; 
\addlegendentry{FI, A1}; 

\end{axis}
\end{tikzpicture}
\caption{Comparison of several plate-tangential adjoint velocity profiles $\hat{v}_\mathrm{1}(x_\mathrm{2})$ for an external (FE, $\hat{v}_\mathrm{1}(0) / \hat{V}_\mathrm{1} = 0$) and internal (FI, $\hat{v}_\mathrm{1}(0) / \hat{V}_\mathrm{1} = -1$) drag formulation. Additionally, several adjoint 99\%-boundary-layer thicknesses $\hat{\delta}_\mathrm{99}$ are depicted, augmented by the expected adjoint Blasius solutions.}
\label{fig:adjoint_velocity_profiles}
\end{figure}
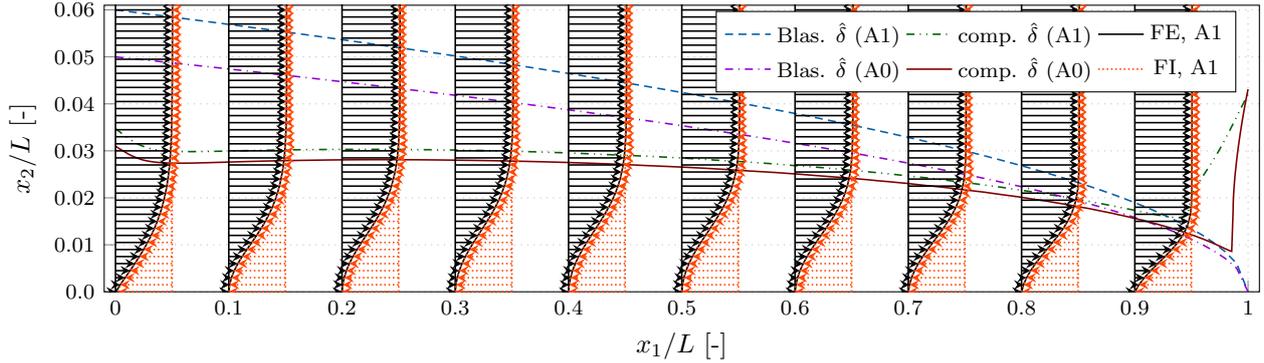

\paragraph{Unified Adjoint Velocity Profile}
As shown in Sec. \ref{sec:continuous_blasius_investigation}, the internal and external adjoint formulation of a force functional can be unified. To illustrate this, results obtained from an exemplary adjoint simulation at $\mathrm{Re}_\mathrm{L}=10^4$ are  discussed in more detail.
Figure \ref{fig:adjoint_velocity_profiles} illustrates predicted adjoint tangential velocity profiles  
at ten equally spaced positions along the plate for both the (FI) as well as the (FE) formulation.
Both sets of exemplary results incorporate ATC (A1). As outlined in Sec. \ref{sec:continuous_blasius_investigation}, the FI and FE adjoint velocity profiles reveal a similarity, since they are based on different formulations that aim at answering the same engineering question. Optically visible and numerically measurable, both adjoint velocity profile sets grow out of the plate with the same gradient. The latter enters the sensitivity along the plate ($x_\mathrm{2} / L = 0$) according to Eqn. (\ref{equ:shape_derivative}).
In addition to the adjoint velocity profiles, the 99\%-b.-l. thickness $\hat{\delta}_\mathrm{99}$ is depicted for various formulations, viz. [FI, FE] $\times$ [A0, A1]. The trend of an increased $\hat{\delta}_\mathrm{99}$ b.-l. thickness due to the influence of ATC (A0 vs. A1) is reproduced by the simulations, though to a lesser extent. 
%

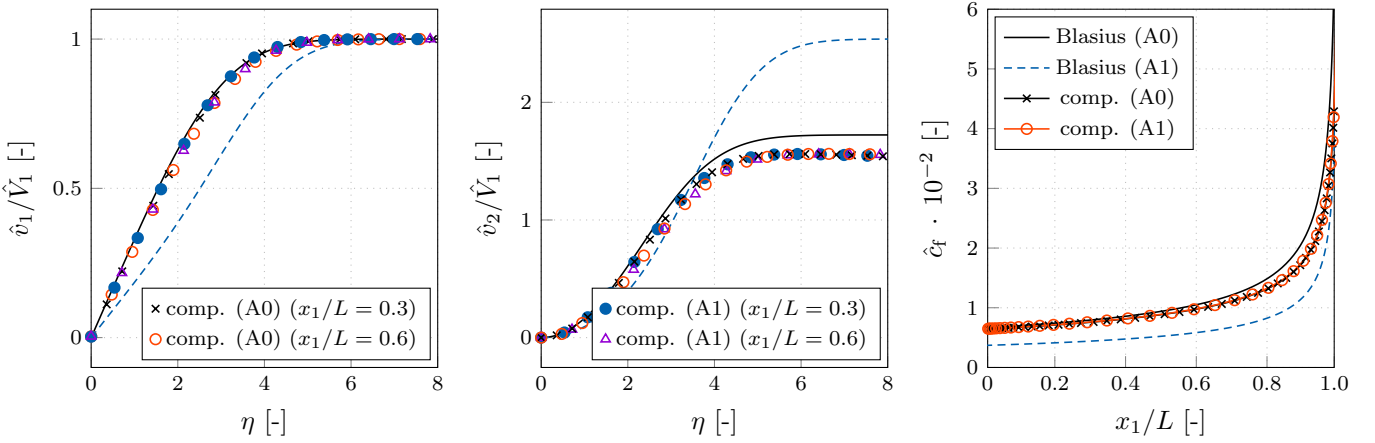
\begin{figure}
\centering
\analytiSolutionPictures
\begin{tikzpicture}
\begin{axis}[
 ylabel style={text width=0.25\textwidth,align=center},
 xlabel={$\eta$ [-]},
 ylabel={$\hat{v}_\mathrm{1} / \hat{V}_\mathrm{1}$ [-]},
 legend style={at={(0.98,0.02)},anchor=south east},
 xmin=0,xmax=8,
]

\addplot [mark1, only marks, each nth point=2] table[x expr={\thisrowno{4}},y expr={\thisrowno{5}}] {data/Plate_Adjoint_Blasius_FreSCo_V1_A0_Results.dat};
\addplot [mark2, only marks, each nth point=2] table[x expr={\thisrowno{10}},y expr={\thisrowno{11}}] {data/Plate_Adjoint_Blasius_FreSCo_V1_A0_Results.dat};
\addplot [mark3, only marks, each nth point=3] table[x expr={\thisrowno{4}},y expr={\thisrowno{5}}] {data/Plate_Adjoint_Blasius_FreSCo_V1_A1_Results.dat};
\addplot [mark4, only marks, each nth point=3] table[x expr={\thisrowno{10}},y expr={\thisrowno{11}}] {data/Plate_Adjoint_Blasius_FreSCo_V1_A1_Results.dat};
\addplot [line1] table[x expr={\thisrowno{0}},y expr={\thisrowno{2}}] {data/Blasius_Shooting_Adjoint_Results_F1_A0.dat};
\addplot [line3] table[x expr={\thisrowno{0}},y expr={\thisrowno{2}}] {data/Blasius_Shooting_Adjoint_Results_F1_A1.dat};

\addlegendentry{comp. (A0) ($x_\mathrm{1}/L = 0.3$)};
\addlegendentry{comp. (A0) ($x_\mathrm{1}/L = 0.6$)};
 
\end{axis}
\end{tikzpicture}
\begin{tikzpicture}
\begin{axis}[
 ylabel style={text width=0.25\textwidth,align=center},
 xlabel={$\eta$ [-]},
 ylabel={$\hat{v}_\mathrm{2} / \hat{V}_\mathrm{1}$ [-]},
 legend style={at={(0.98,0.02)},anchor=south east},
 xmin=0,xmax=8,
]

\addplot [mark3, only marks, each nth point=3] table[x expr={\thisrowno{4}},y expr={\thisrowno{5}}] {data/Plate_Adjoint_Blasius_FreSCo_V2_A1_Results.dat};
\addplot [mark4, only marks, each nth point=3] table[x expr={\thisrowno{10}},y expr={\thisrowno{11}}] {data/Plate_Adjoint_Blasius_FreSCo_V2_A1_Results.dat};
\addplot [mark1, only marks, each nth point=2] table[x expr={\thisrowno{4}},y expr={\thisrowno{5}}] {data/Plate_Adjoint_Blasius_FreSCo_V2_A0_Results.dat};
\addplot [mark2, only marks, each nth point=2] table[x expr={\thisrowno{10}},y expr={\thisrowno{11}}] {data/Plate_Adjoint_Blasius_FreSCo_V2_A0_Results.dat};
\addplot [line1] table[x expr={\thisrowno{0}},y expr={\thisrowno{4}}] {data/Blasius_Shooting_Adjoint_Results_F1_A0.dat};
\addplot [line3] table[x expr={\thisrowno{0}},y expr={\thisrowno{4}}] {data/Blasius_Shooting_Adjoint_Results_F1_A1.dat};

\addlegendentry{comp. (A1) ($x_\mathrm{1}/L = 0.3$)};
\addlegendentry{comp. (A1) ($x_\mathrm{1}/L = 0.6$)};
 
\end{axis}
\end{tikzpicture}
\begin{tikzpicture}
\begin{axis}[
 ylabel style={text width=0.25\textwidth,align=center},
 xlabel={$x_\mathrm{1} / L$ [-]},
 ylabel={$\hat{c}_\mathrm{f} \cdot 10^{-2}$ [-]},
 legend style={at={(0.02,0.98)},anchor=north west},
 xmin=0.01,xmax=0.99,
 ymin=0.0,ymax=0.06,
 xtick={0.01,0.2,0.4,0.6,0.8,0.99},
 xticklabels={0,0.2,0.4,0.6,0.8,1.0},
 ytick={0.0,0.01,0.02,0.03,0.04,0.05,0.06},
 yticklabels={0,1,2,3,4,5,6},
 scaled y ticks = false
]
\addplot [line1] table[x expr={\thisrowno{0}},y expr={\thisrowno{4}}] {data/Blasius_Shooting_Adjoint_Friction_Coefficient.dat};
\addplot [line3] table[x expr={\thisrowno{0}},y expr={\thisrowno{2}}] {data/Blasius_Shooting_Adjoint_Friction_Coefficient.dat};
\addplot [mark1, each nth point=2] table[x expr={\thisrowno{0}},y expr={\thisrowno{3}}] {data/Blasius_Shooting_Adjoint_Friction_Coefficient.dat};
\addplot [mark2, each nth point=3] table[x expr={\thisrowno{0}},y expr={\thisrowno{1}}] {data/Blasius_Shooting_Adjoint_Friction_Coefficient.dat};
 
\addlegendentry{Blasius (A0)};
\addlegendentry{Blasius (A1)};
\addlegendentry{comp. (A0)};
\addlegendentry{comp. (A1)};

\end{axis}
\end{tikzpicture}
\caption{Local results for the adjoint flow over a flat plate at $\mathrm{Re}_\mathrm{L} = 10^4$ based on a manipulated primal field: Left) tangential adjoint velocity ($\hat{v}_\mathrm{1} / \hat{V}_\mathrm{1}$) against $\hat{f}^\prime$ and center) normal adjoint velocity ($\hat{v}_\mathrm{2} / \hat{V}_\mathrm{1}$) against $\hat{f}^\prime \hat{\eta} - \hat{f}$ both over the similarity variable. Right) adjoint friction coefficient $\hat{c}_\mathrm{f}$ against the adjoint Blasius estimations based on an interface measure that employs $a=L$ and $b=-1$.}
\label{fig:plate_adjoint_local_results}
\end{figure}

\paragraph{Manipulated Primal Field}
The investigated boundary-layer contradicts the assumption of an infinitely long plate immanent to the theory, and  deviations between the computed Navier-Stokes results and the Blasius solution increase towards the trailing edge of the plate. However, due to the (approximately) parabolic nature of the computed flow, deviations are primarily transported downstream and their significant local influence documented by Fig. \ref{fig:plate_primal_local_results} has hardly any effect on an integral level, cf. Fig. \ref{fig:plate_primal_integral_results}.
A different picture emerges for the adjoint flow. 
Since the adjoint solution propagates from the trailing to the leading edge, the processes reverse their direction in adjoint mode and primal flow deviations are introduced at the adjoint upstream location. Therefore, the comparison of computed and Blasius adjoint results for a plate of finite length is afflicted by initial value deviations of the primal field and might be debatable. For this reason, the primal velocity field at $\mathrm{Re}_\mathrm{L} = 10^4$ was manipulated by re-initialising a velocity field that follows from the similarity transformation of a computed primal velocity profile extracted at x/L = 3/4.  
%
Subsequently all four adjoint simulations were performed based on  $a=L$ and $b=-1$. 
Figure \ref{fig:plate_adjoint_local_results}  outlines the computed results for the normalized adjoint tangential $\hat{v}_\mathrm{1} / \hat{V}_\mathrm{1}$ (left) and normal  $\hat{v}_\mathrm{2} / \hat{V}_\mathrm{1}$ (center) velocity
next to the adjoint friction coefficient (right) in comparison to the Blasius solutions with (A1) and without (A0) ATC. 
%
Only minor deviations are observed when comparing the Navier-Stokes predictions for different adjoint formulations. However, an improved agreement with the Blasius solution that neglects the ATC (A0) is clearly seen. This supports the results of the similarity transformation, i.e. Eqn. (\ref{equ:tangential_adjoint_blasius_equation})-(\ref{equ:normal_adjoint_blasius_equation}), according to which ATC influence should disappear in laminar flat plate b.-l. flows. Analogue to the primal investigation, the plate tangential adjoint velocity profiles are closer to their companion Blasius solutions than the normal velocity profiles. However, a smaller normal velocity variation is observed between the different locations. Moreover the manipulation of the (upstream) primal results reveals benefits for the agreement between the Blasius and the computed adjoint skin friction. 

\begin{figure}
\centering
\analytiSolutionPictures
\begin{tikzpicture}
\begin{axis}[
 ylabel style={text width=0.25\textwidth,align=center},
 legend style={at={(0.98,0.02)},anchor=south east},
 xlabel={$\mathrm{Re}_\mathrm{L} \cdot 10^5$ [-]},
 ylabel={$\hat{c}_\mathrm{d} \cdot 10^{-2}$ [-]},
 xmin=1000,xmax=100000,
 ymin=-0.04,ymax=0.0,
 xtick={1000,20000,40000,60000,80000,100000},
 xticklabels={0.01,0.2,0.4,0.6,0.8,1.0},
 scaled x ticks = false,
 ytick={-0.04,-0.03,-0.02,-0.01,0.0},
 yticklabels={-4,-3,-2,-1,0.0},
 scaled y ticks = false
]
\addplot [line1] table[x expr={\thisrowno{0}},y expr={\thisrowno{2}}] {data/Plate_Adjoint_Drag_Coefficient.dat};
\addplot [line3] table[x expr={\thisrowno{0}},y expr={\thisrowno{1}}] {data/Plate_Adjoint_Drag_Coefficient.dat};
\addplot [line4, mark1] table[x expr={\thisrowno{0}},y expr={\thisrowno{4}}] {data/Plate_Adjoint_Drag_Coefficient.dat};
\addplot [line5, mark2] table[x expr={\thisrowno{0}},y expr={\thisrowno{3}}] {data/Plate_Adjoint_Drag_Coefficient.dat};
 
\addlegendentry{Blasius (A0)};
\addlegendentry{Blasius (A1)};
\addlegendentry{comp. (A0)};
\addlegendentry{comp. (A1)};
\end{axis}
\end{tikzpicture}
\begin{tikzpicture}
\begin{axis}[
 ylabel style={text width=0.25\textwidth,align=center},
 xlabel={$\mathrm{Re}_\mathrm{L} \cdot 10^5$ [-]},
 ylabel={$\hat{\delta}_\mathrm{99} / (3L/4) \cdot 10^{-2}$ [-]},
 xmin=1000,xmax=100000,
 ymin=0.0,ymax=0.06,
 xtick={1000,20000,40000,60000,80000,100000},
 xticklabels={0.01,0.2,0.4,0.6,0.8,1.0},
 scaled x ticks = false,
 ytick={0.0,0.01,0.02,0.03,0.04,0.05,0.06},
 yticklabels={0,1,2,3,4,5,6},
 scaled y ticks = false
]
\addplot [line1] table[x expr={\thisrowno{0}},y expr={\thisrowno{2}}] {data/Plate_Adjoint_Integral_Thicknesses.dat};
\addplot [line3] table[x expr={\thisrowno{0}},y expr={\thisrowno{1}}] {data/Plate_Adjoint_Integral_Thicknesses.dat};
\addplot [line4, mark1] table[x expr={\thisrowno{0}},y expr={\thisrowno{4}}] {data/Plate_Adjoint_Integral_Thicknesses.dat};
\addplot [line5, mark2] table[x expr={\thisrowno{0}},y expr={\thisrowno{3}}] {data/Plate_Adjoint_Integral_Thicknesses.dat};
 
\addlegendentry{Blasius (A0)};
\addlegendentry{Blasius (A1)};
\addlegendentry{comp. (A0)};
\addlegendentry{comp. (A1)};

\end{axis}
\end{tikzpicture}
\begin{tikzpicture}
\begin{axis}[
 ylabel style={text width=0.25\textwidth,align=center},
 xlabel={$\mathrm{Re}_\mathrm{L} \cdot 10^5$ [-]},
 legend style={at={(0.98,0.02)},anchor=south east},
 ylabel={$c_\mathrm{\delta_\mathrm{u} J}$ [-]},
 xmin=1000,xmax=50000,
 ymin=-0.75,ymax=-0.5,
 xtick={1000,20000,40000,60000,80000,100000},
 xticklabels={0.01,0.2,0.4,0.6,0.8,1.0},
 scaled x ticks = false
]
\addplot [line5] table[x expr={\thisrowno{0}},y expr={\thisrowno{1}}] {data/Plate_Adjoint_Integral_Sensitivity_Dummy.dat};
\addplot [line6] table[x expr={\thisrowno{0}},y expr={\thisrowno{2}}] {data/Plate_Adjoint_Integral_Sensitivity_Dummy.dat};
\addplot [line4, mark1] table[x expr={\thisrowno{0}},y expr={\thisrowno{2}}] {data/Plate_Adjoint_Integral_Sensitivity.dat};
\addplot [line5, mark2] table[x expr={\thisrowno{0}},y expr={\thisrowno{1}}] {data/Plate_Adjoint_Integral_Sensitivity.dat};
 
\addlegendentry{$\mathcal{O}(- \mathrm{Re}_\mathrm{L}^{-1/2})$};
\addlegendentry{$\mathcal{O}(- \mathrm{Re}_\mathrm{L}^{-1})$};
\addlegendentry{comp. (A0)};
\addlegendentry{comp. (A1)};

\end{axis}
\end{tikzpicture}
\caption{Integral results for the adjoint flow over a flat plate: Left) adjoint drag coefficient $\hat{c}_\mathrm{d}$, center) adjoint boundary-layer thickness ($\hat{\delta}_\mathrm{99}$) at $x/L = 1/4$ and right) integral shape sensitivity coefficient $c_\mathrm{\delta_\mathrm{u} J}$ over the plate-length based Reynolds-number $\mathrm{Re}_\mathrm{L}$. }
\label{fig:plate_adjoint_integral_results}
\end{figure}
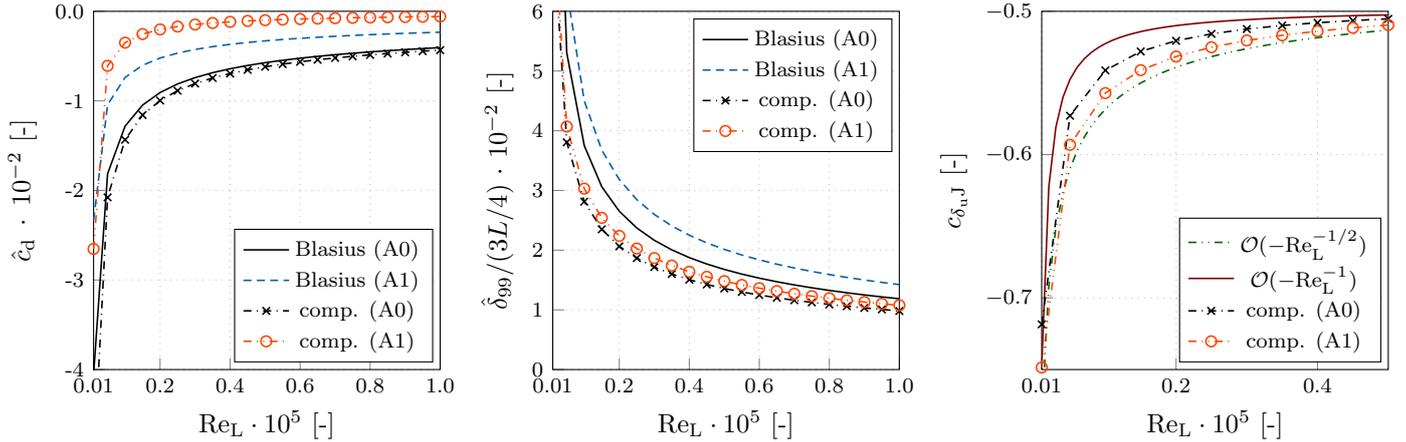

\paragraph{Consistent Primal Field}
The final subsection discusses computed adjoint Navier-Stokes results based upon non-manipulated predictions of the primal flow, which are displayed in Figs. \ref{fig:plate_primal_integral_results}-\ref{fig:plate_primal_local_results}. 
%
Figure \ref{fig:plate_adjoint_integral_results} shows the evolution of the adjoint drag-coefficient $\hat{c}_\mathrm{d}$ (left) and the adjoint b.-l. thickness $\hat{\delta}_\mathrm{99}$ (center) over the Reynolds number for the two considered adjoint formulations (A0, A1). The adjoint b.-l. thickness is evaluated at $x_\mathrm{1} / L= 1/4$ based on a measure that employs $ a= L$ and $b = -1$.
Regarding the adjoint drag coefficient, the predicted results of the formulation without ATC (A0) are again significantly closer to the corresponding Blasius results than those that include ATC (A1).
As already shown in Fig. \ref{fig:adjoint_velocity_profiles} for $\mathrm{Re}_\mathrm{L} = 10^4$, simulations without adjoint advection result in a reduced b.-l. thickness in line with the observations of Sec. \ref{sec:numerical_blasius_approximation}.
The investigations in Sec. \ref{sec:continuous_blasius_investigation} reveal that the integral shape sensitivity  scales with $1/\mathrm{Re}_\mathrm{L}$ but approaches a singularity towards the leading edge of the plate. Therefore, no adjoint Blasius solutions are shown in Fig. \ref{fig:plate_adjoint_integral_results} (right).  Instead, the numerical results expressed by the dimensionless sensitivity coefficient are compared with two reference curves, which scale with $1/\mathrm{Re}_\mathrm{L}$ and $1/\sqrt{\mathrm{Re}_\mathrm{L}}$ respectively.
The computed Navier-Stokes results fall in between these two reference curves, but tend to follow the inverse root relation.
Due to the more singular behavior in comparison to e.g. the adjoint drag coefficient ($\mathcal{O}(- \mathrm{Re}_\mathrm{L}^{-1})$ vs. $\mathcal{O}(- \mathrm{Re}_\mathrm{L}^{-1/2})$), only the range between $10^3 \leq \mathrm{Re}_\mathrm{L} \leq 5 \cdot 10^4$ is shown.
We conclude: Consistent, real (finite-length) plate investigations w.r.t. integral adjoint quantities are qualitatively in very good agreement with analytic estimations from Sec. \ref{sec:continuous_blasius_investigation}. The quantitative deterioration follows mainly from the already quantitatively incorrect primal results (cf. Fig. \ref{fig:plate_primal_integral_results} - \ref{fig:plate_primal_local_results}).

%
\section{Conclusions and Outlook}
\label{sec:conclusion}
The paper reports the derivation and analysis of a continuous adjoint complement to the Blasius equation for a flat plate boundary-layer. The discussion is divided into three main sections and arrives at two major conclusions.

The first part demonstrated two alternatives for deriving the adjoint boundary-layer equations, either following a \textit{simplify-and-derive} or a \textit{derive-and-simplify} strategy. The procedure reminds of two prominent classical adjoint strategies, frequently subdivided into discrete adjoint (\textit{discretize-and-derive}) vs. continuous adjoint (\textit{derive-and-discretize}).

The second part was devoted to the analysis of the coupled primal/adjoint b.-l. framework and led to generalized similarity parameters. A first major conclusion refers to the existence of virtually identical similarity parameters to condense the primal and the adjoint b.-l. flows. The latter turn the  Partial-Differential-Equation problem into a boundary value problem described by a set of Ordinary-Differential-Equations and support the formulation of an adjoint complement to the classical Blasius equation. Opposite to the primal Blasius equation, its adjoint complement consists of two ODEs which can be simplified depending on the treatment of adjoint advection also known as adjoint transpose convection. It was shown, that the advective fluxes vanish for the investigated self-similar b.-l. flows. This second major conclusion supports the heuristic neglect of the term used by many authors of continuous adjoint optimization studies into complex engineering shear flows.
Moreover, a formalism was derived for analytical expressions of an adjoint b.-l. thickness, wall shear stress, skin friction and drag coefficient as well as a shape sensitivity expression for a shear driven drag objective.

The third part assessed the predictive agreement between the different Blasius solutions and numerical results for Navier-Stokes simulations of a flat plate b.-l. at Reynolds numbers between $10^3 \leq \mathrm{Re}_\mathrm{L} \leq 10^5$. The reversal of the inlet and outlet location as well as the direction of the flow, inherent to the adjoint formulation of convective kinematics, poses a challenge when investigating real finite length
flat plate boundary layer flows. The neglect of the adjoint transposed convection term was confirmed by the NS results and the predicted  adjoint integral quantities 
agree with adjoint Blasius results.

Future work should investigate other self-similar flows, e.g. derive adjoint Falkner-Skan solutions for wedged geometries and develop an adjoint solution for compressible boundary layers.
Volume-based objective functionals could be investigated, e.g. to account for variations in the primal flow field that potentially build a bridge towards the stability crisis immanent to flow transition. 
Finally, research towards turbulent b.-l. flows might be fruitful, whereby one could follow both the continuous as well as the discrete adjoint approach in parallel.

\section{Acknowledgments}
The current work is a part of the research projects "Drag Optimisation of Ship Shapes’" funded by the German Research Foundation (DFG, Grant No. RU 1575/3-1)
as well as "Dynamic Adaptation of Modular Shape Optimization Processes" funded by the German Federal Ministry for Economic Affairs and Energy (BMWi, Grant No.  03SX453B).
Moreover, the research takes places within the Research Training Group (RTG) 2583 "Modeling, Simulation and Optimization with Fluid Dynamic Applications" funded by the German Research Foundation.
This support is gratefully acknowledged by the authors. 

\section{Authorship Contribution Statement}
\textbf{Niklas K{\"u}hl}: Conceptualization, Methodology, Software, Validation, Formal analysis, Investigation, Writing - original draft, Visualization, Writing - review \& editing.
\textbf{Peter M. M\"uller} : Methodology, Formal analysis, Writing - review \& editing.
\textbf{Thomas Rung}: Project administration, Funding acquisition, Supervision, Resources,  Writing - review \& editing.

\bibliographystyle{plain}
\bibliography{./library.bib}

\begin{appendix}
\section{Primal Scaling Analysis}
\label{app:primal_simplification}
Performing a non-dimensionalization of Eqn. (\ref{equ:primal_momentum})-(\ref{equ:primal_mass}) with the reference data given in Tab. (\ref{tab:reference_values}) results in:
\begin{alignat}{2}
R_\mathrm{1}^*:& 
v_\mathrm{1}^* \frac{\partial v_\mathrm{1}^*}{\partial x_\mathrm{1}^*} \frac{V_\mathrm{1} V_\mathrm{1}}{L} 
+ v_\mathrm{2}^* \frac{\partial v_\mathrm{1}^*}{\partial x_\mathrm{2}^*}\frac{V_\mathrm{2} V_\mathrm{1}}{\delta} 
+ \frac{P}{L \, \rho} \frac{\partial p^*}{\partial x_\mathrm{1}^*}  
- \nu  \frac{V_\mathrm{1}}{L^2} \left[ \frac{\partial^2 v_\mathrm{1}^*}{\partial x_\mathrm{1}^{*2}} -  \frac{L^2}{\delta^2} \frac{\partial^2 v_\mathrm{1}^*}{\partial x_\mathrm{2}^{*2}} \right] 
&&= 0 \\
R_\mathrm{2}^*:& 
v_\mathrm{1}^* \frac{\partial v_\mathrm{2}^*}{\partial x_\mathrm{1}^*} \frac{V_\mathrm{1} V_\mathrm{2}}{L} 
+ v_\mathrm{2}^* \frac{\partial v_\mathrm{2}^*}{\partial x_\mathrm{2}^*} \frac{V_\mathrm{2} V_\mathrm{2}}{\delta} 
+ \frac{P}{\delta \rho} \frac{\partial p^*}{\partial x_\mathrm{2}^*}
- \nu \frac{V_\mathrm{2}}{L^2} \left[ \frac{\partial^2 v_\mathrm{2}^*}{\partial x_\mathrm{1}^{*2}} +  \frac{L^2}{\delta^2} \frac{\partial^2 v_\mathrm{2}^*}{\partial x_\mathrm{2}^{*2}} \right] 
&&= 0 \\
Q^*:& 
-\frac{\partial v_\mathrm{1}^*}{\partial x_\mathrm{1}^*} \frac{V_\mathrm{1}}{L}
-\frac{\partial v_\mathrm{2}^*}{\partial x_\mathrm{2}^*} \frac{V_\mathrm{2}}{\delta}
&&= 0.
\end{alignat}
that can be simplified towards:
\begin{alignat}{2}
R_\mathrm{1}^*:& 
v_\mathrm{1}^* \frac{\partial v_\mathrm{1}^*}{\partial x_\mathrm{1}^*}
+ v_\mathrm{2}^* \frac{\partial v_\mathrm{1}^*}{\partial x_\mathrm{2}^*}
+ \mathrm{Eu} \frac{\partial p^*}{\partial x_\mathrm{1}^*}
- \frac{1}{\mathrm{Re}_\mathrm{L}} \left[ \frac{\partial^2 v_\mathrm{1}^*}{\partial x_\mathrm{1}^{*2}} +  \frac{\partial^2 v_\mathrm{1}^*}{\partial x_\mathrm{2}^{*2}} \frac{L^2}{\delta^2} \right] 
&&= 0 \\
R_\mathrm{2}^*:& 
v_\mathrm{1}^* \frac{\partial v_\mathrm{2}^*}{\partial x_\mathrm{1}^*}
+ v_\mathrm{2}^* \frac{\partial v_\mathrm{2}^*}{\partial x_\mathrm{2}^*} 
+ \mathrm{Eu} \frac{L^2}{\delta^2} \frac{\partial p^*}{\partial x_\mathrm{2}^*} 
- \frac{1}{\mathrm{Re}_\mathrm{L}} \left[ \frac{\partial^2 v_\mathrm{2}^*}{\partial x_\mathrm{1}^{*2}} + \frac{\partial^2 v_\mathrm{2}^*}{\partial x_\mathrm{2}^{*2}} \frac{L^2}{\delta^2} \right] 
&&= 0 \\
Q^*:& 
-\frac{\partial v_\mathrm{1}^*}{\partial x_\mathrm{1}^*} 
-\frac{\partial v_\mathrm{2}^*}{\partial x_\mathrm{2}^*} 
&&= 0.
\end{alignat}
by assuming $V_\mathrm{2} \propto V_\mathrm{1} \delta / L$ and defining the Reynolds- and Euler number as $\mathrm{Re}_\mathrm{L} = \frac{L V_\mathrm{1}}{\nu}$ and $\mathrm{Eu} = \frac{P}{\rho V_\mathrm{1}^2}$ respectively.

\section{Adjoint Scaling Analysis}
\label{app:adjoint_simplification}
Performing a non-dimensionalization of Eqn. (\ref{equ:adjoint_momentum})-(\ref{equ:adjoint_mass}) with the reference data given in Tab. (\ref{tab:reference_values}) results in:
\begin{alignat}{2}
\hat{R}_\mathrm{1}^*:&
- v_\mathrm{1}^* \frac{\partial \hat{v}_\mathrm{1}^*}{\partial x_\mathrm{1}^*} \frac{V_\mathrm{1} \hat{V}_\mathrm{1}}{L}
- v_\mathrm{2}^* \frac{\partial \hat{v}_\mathrm{1}^*}{\partial x_\mathrm{2}^*} \frac{V_\mathrm{2} \hat{V}_\mathrm{1}}{\hat{\delta}}
+ \hat{v}_\mathrm{1}^* \frac{\partial v_\mathrm{1}^*}{\partial x_\mathrm{1}^*} \frac{\hat{V}_\mathrm{1} V_\mathrm{1}}{L}
+ \hat{v}_\mathrm{2}^* \frac{\partial v_\mathrm{2}^*}{\partial x_\mathrm{1}^*} \frac{\hat{V}_\mathrm{2} V_\mathrm{2}}{L}
+ \frac{\hat{P}}{L} \frac{\partial \hat{p}^*}{\partial x_\mathrm{1}^*}
- \nu \frac{\hat{V}_\mathrm{1}}{L^2}  \left[ \frac{\partial^2 \hat{v}_\mathrm{1}^*}{\partial x_\mathrm{1}^{*2}} + \frac{L}{\hat{\delta}}  \frac{\partial^2 \hat{v}_\mathrm{1}^*}{\partial x_\mathrm{2}^{*2}} \right]
&&= -\frac{\partial j_\mathrm{\Omega}^*}{\partial v_\mathrm{1}^*} \frac{J_\mathrm{\Omega}}{V_\mathrm{1}} \\
\hat{R}_\mathrm{2}^*:&
- v_\mathrm{1}^* \frac{\partial \hat{v}_\mathrm{2}^*}{\partial x_\mathrm{1}^*} \frac{V_\mathrm{1} \hat{V}_\mathrm{2}}{L}
- v_\mathrm{2}^* \frac{\partial \hat{v}_\mathrm{2}^*}{\partial x_\mathrm{2}^*} \frac{V_\mathrm{2} \hat{V}_\mathrm{2}}{\hat{\delta}}
+ \hat{v}_\mathrm{1}^* \frac{\partial v_\mathrm{1}^*}{\partial x_\mathrm{2}^*} \frac{\hat{V}_\mathrm{1} V_\mathrm{1}}{\delta}
+ \hat{v}_\mathrm{2}^* \frac{\partial v_\mathrm{2}^*}{\partial x_\mathrm{2}^*} \frac{\hat{V}_\mathrm{2} V_\mathrm{2}}{\delta}
+ \frac{\hat{P}}{\hat{\delta}} \frac{\partial \hat{p}^*}{\partial x_\mathrm{2}^*}
- \nu \frac{\hat{V}_\mathrm{2}}{L^2} \left[ \frac{\partial^2 \hat{v}_\mathrm{2}^*}{\partial x_\mathrm{1}^{*2}} + \frac{L}{\hat{\delta}} \frac{\partial^2 \hat{v}_\mathrm{2}^*}{\partial x_\mathrm{2}^{*2}} \right]
&&=  -\frac{\partial j_\mathrm{\Omega}^*}{\partial v_\mathrm{2}^*} \frac{J_\mathrm{\Omega}}{V_\mathrm{2}}  \\
\hat{Q}^*:&  
-\frac{1}{\rho}\frac{\partial \hat{v}_\mathrm{1}^*}{\partial x_\mathrm{1}^*} \frac{\hat{V}_\mathrm{1}}{L} 
-\frac{1}{\rho}\frac{\partial \hat{v}_\mathrm{2}^*}{\partial x_\mathrm{2}^*} \frac{\hat{V}_\mathrm{2}}{\hat{\delta}}
&&=0
\end{alignat}
that can be simplified towards:
\begin{alignat}{2}
\hat{R}_\mathrm{1}^*:&
- v_\mathrm{1}^* \frac{\partial \hat{v}_\mathrm{1}^*}{\partial x_\mathrm{1}^*}
- v_\mathrm{2}^* \frac{\partial \hat{v}_\mathrm{1}^*}{\partial x_\mathrm{2}^*} \frac{\delta}{\hat{\delta}}
+ \hat{v}_\mathrm{1}^* \frac{\partial v_\mathrm{1}^*}{\partial x_\mathrm{1}^*}
+ \hat{v}_\mathrm{2}^* \frac{\partial v_\mathrm{2}^*}{\partial x_\mathrm{1}^*} \frac{\hat{\delta}^2}{L^2}
+ \frac{\hat{P}}{V_\mathrm{1} \hat{V}_\mathrm{1}} \frac{\partial \hat{p}^* }{\partial x_\mathrm{1}^*}
- \frac{1}{\mathrm{Re}_\mathrm{L}} \left[ \frac{\partial^2 \hat{v}_\mathrm{1}^*}{\partial x_\mathrm{1}^{*2}} + \frac{\partial^2 \hat{v}_\mathrm{1}^*}{\partial x_\mathrm{2}^{*2}} \frac{L^2}{\hat{\delta}^2} \right]
&&= -\frac{\partial j_\mathrm{\Omega}^*}{\partial v_\mathrm{1}^*} \frac{J_\mathrm{\Omega} L}{V_\mathrm{1}^2 \hat{V}_\mathrm{1}} \\
\hat{R}_\mathrm{2}^*:&
- v_\mathrm{1}^* \frac{\partial \hat{v}_\mathrm{2}^*}{\partial x_\mathrm{1}^*} 
- v_\mathrm{2}^* \frac{\partial \hat{v}_\mathrm{2}^*}{\partial x_\mathrm{2}^*} \frac{\delta}{\hat{\delta}} 
+ \hat{v}_\mathrm{1}^* \frac{\partial v_\mathrm{1}^*}{\partial x_\mathrm{2}^*} \frac{L^2}{\delta \hat{\delta}}
+ \hat{v}_\mathrm{2}^* \frac{\partial v_\mathrm{2}^*}{\partial x_\mathrm{2}^*} 
+ \frac{\partial \hat{p}^*}{\partial x_\mathrm{2}^*} \frac{\hat{P}}{\hat{V}_\mathrm{1} V_\mathrm{1}} \frac{L^2}{\hat{\delta}^2}
- \frac{1}{\mathrm{Re}_\mathrm{L}} \left[ \frac{\partial^2 \hat{v}_\mathrm{2}^*}{\partial x_\mathrm{1}^{*2}}  +  \frac{L^2}{\hat{\delta}^2} \frac{\partial^2 \hat{v}_\mathrm{2}^{*}}{\partial x_\mathrm{2}^{*2}} \right]
&&= -\frac{\partial j_\mathrm{\Omega}^*}{\partial v_\mathrm{2}^*} \frac{J_\mathrm{\Omega} L^3}{V_\mathrm{1}^2 \hat{V}_\mathrm{1} \delta \hat{\delta}}\\
\hat{Q}^*:&
- \frac{\partial \hat{v}_\mathrm{1}^*}{\partial x_\mathrm{1}^*} - \frac{\partial \hat{v}_\mathrm{2}^*}{\partial x_\mathrm{2}^*}
&&= 0
\end{alignat}
by assuming $\hat{V}_\mathrm{2} \propto \hat{V}_\mathrm{1} \hat{\delta} / L$.

\section{Primal and Adjoint Similarity Relations}
\label{app:similarity_relations}

The relations in Tab. (\ref{tab:similarity_relations}) simplify the tangential primal and the tangential as well as normal adjoint b.-l. equations.
\begin{table}
\centering
\begin{tabular}{|c||c|c|c|c|c|}
\hline
primal quantity/operator & 
$v_\mathrm{1}$  & 
$v_\mathrm{2}$  & 
$\partial v_\mathrm{1} / \partial x_\mathrm{1}$ & 
$\partial v_\mathrm{1} / \partial x_\mathrm{2}$ & 
$\partial^2 v_\mathrm{1} / \partial x_\mathrm{2}^2$ \\
\hline
similarity transformation & 
$V_\mathrm{1} f^{\prime}$ & 
$\frac{1}{2} \sqrt{\frac{\nu b^2 V_\mathrm{1}}{a + b x_\mathrm{1}}} \left[ f^{\prime} \eta -f \right]$ & 
$- \frac{1}{2} \frac{V_\mathrm{1} b}{a + b x_\mathrm{1}} f^{\prime \prime} \eta$ & 
$V_\mathrm{1} \sqrt{\frac{V_\mathrm{1}}{ \nu (a + b x_\mathrm{1})}} f^{\prime \prime} $ & 
$V_\mathrm{1} \frac{V_\mathrm{1}}{ \nu (a + b x_\mathrm{1})} f^{\prime \prime \prime} $ \\
\hline
\hline
adjoint quantity/operator & 
$\hat{v}_\mathrm{1}$  & 
$\hat{v}_\mathrm{2}$  & 
$\partial \hat{v}_\mathrm{1} / \partial x_\mathrm{1}$ & 
$\partial \hat{v}_\mathrm{1} / \partial x_\mathrm{2}$ & 
$\partial^2 \hat{v}_\mathrm{1} / \partial x_\mathrm{2}^2$ \\
\hline
similarity transformation & 
$\hat{V}_\mathrm{1} \hat{f}^{\prime}$ & 
$\frac{1}{2} \sqrt{\frac{\nu b^2 V_\mathrm{1}}{a + b x_\mathrm{1}}} \left[ \hat{f}^{\prime} \hat{\eta} -\hat{f} \right]$ & 
$- \frac{1}{2} \frac{\hat{V}_\mathrm{1} b}{a + b x_\mathrm{1}} \hat{f}^{\prime \prime} \hat{\eta}$ & 
$\hat{V}_\mathrm{1} \sqrt{\frac{V_\mathrm{1}}{ \nu (a + b x_\mathrm{1})}} \hat{f}^{\prime \prime} $ & 
$\hat{V}_\mathrm{1} \frac{V_\mathrm{1}}{ \nu (a + b x_\mathrm{1})} \hat{f}^{\prime \prime \prime} $ \\
\hline
\end{tabular}
\caption{Similarity relations for the primal and the adjoint boundary-layer equations.}
\label{tab:similarity_relations}
\end{table}

\end{appendix}

\end{document}